\documentclass[10pt]{article}
\usepackage{amssymb}
\usepackage{stmaryrd}
\usepackage{amsthm}
\usepackage{mathpartir}
\usepackage{times}
\usepackage{url}

\newtheorem{definition}{Definition}
\newtheorem{theorem}{Theorem}
\newtheorem{lemma}[theorem]{Lemma}
\newtheorem{corollary}[theorem]{Corollary}
\newtheorem{proposition}[theorem]{Proposition}

\newcounter{ctr:list1}
\newcounter{ctr:list2}

\newenvironment{list2'}
    {\setcounter{ctr:list2}{0}
    \begin{list}{(\roman{ctr:list2})}{\usecounter{ctr:list2}\setlength{\leftmargin}{0.3cm}}}
    {\end{list}}

\newenvironment{mylist}
    {\begin{list}{$\bullet$}{\setlength{\leftmargin}{0.5cm}}}
    {\end{list}}

\newcounter{Crules}
\newcounter{Cassums}

\newcommand{\In}{\mathtt{in}}
\newcommand{\func}[1]{{\bf #1}}
\newcommand{\clk}{\mathsf{Clk}}
\newcommand{\op}{{\it op}}
\newcommand{\adm}{{\it adm}}
\newcommand{\tup}[1]{\langle #1\rangle}
\newcommand{\pause}{;}
\newcommand{\snd}[3]{\overline{#1}\tup{#2}\pause #3}

\newcommand{\rcv}[3]{#1(#2)\pause #3}

\newcommand{\new}[2]{(\nu #1)\:#2}

\newcommand{\cond}[4]{\mathsf{if}\:#1=#2\:\mathsf{then}\:#3\:\mathsf{else}\:#4}

\newcommand{\Req}{\mathsf{Req}}

\newcommand{\Comm}{\mathsf{EOk}}
\newcommand{\PReq}{\mathsf{AReq}}
\newcommand{\CReq}{\mathsf{TReq}}

\newcommand{\AReq}{\mathsf{CReq}}

\newcommand{\DReq}{\mathsf{DReq}}
\newcommand{\action}[1]{\stackrel{#1}{\longrightarrow}}
\newcommand{\fn}{\mathtt{fn}}
\newcommand{\fv}{\mathtt{fv}}
\newcommand{\fnv}{\mathtt{fnv}}

\newcommand{\inferc}[1]{\infer[(#1)]}
\newcommand{\dom}{\mathtt{dom}}

\newcommand{\seq}{\widetilde}

\newcommand{\betac}{\beta^\circ}
\newcommand{\alphac}{\alpha^\circ}
\newcommand{\gammac}{\gamma^\circ}
\newcommand{\deltac}{\delta^\circ}

\newcommand{\subs}[2]{\{{}^{#1}\!/\!{}_{#2}\}}

\newcommand{\pisymbol}{\sigma}

\newcommand{\longrightleftarrows}{\overrightarrow\longleftarrow~~}

\makeatletter
\newcommand{\R}{\addtocounter{Crules}{1}R\arabic{Crules}\gdef\@currentlabel{\arabic{Crules}}}
\newcommand{\A}{\addtocounter{Cassums}{1}A\arabic{Cassums}\gdef\@currentlabel{\arabic{Cassums}}}
\makeatother

\title{On Secure Distributed Implementations of Dynamic Access Control}
\author{Avik Chaudhuri \\ {\normalsize University of California at Santa Cruz} \\ {\normalsize \tt avik@cs.ucsc.edu}}
\date{}

\begin{document}
\maketitle

\begin{abstract} 
\noindent
Distributed implementations of access control abound in distributed storage protocols. While such implementations are often accompanied by informal justifications of their correctness, our formal analysis reveals that their correctness can be tricky. In particular, we discover several subtleties in a state-of-the-art implementation based on capabilities, that can undermine correctness under a simple specification of access control. 

We consider both safety and security for correctness; loosely, safety requires that an implementation does not introduce unspecified behaviors, and security requires that an implementation preserves the specified behavioral equivalences. 
We show that a secure implementation of a static access policy already requires some care in order to prevent unspecified leaks of information about the access policy. 
A dynamic access policy causes further problems. For instance, if accesses can be dynamically granted then the implementation does not remain secure---it leaks information about the access policy. If accesses can be dynamically revoked then the implementation does not even remain safe. We show that a safe implementation is possible if a clock is introduced in the implementation. A secure implementation is possible if the specification is accordingly generalized.

Our analysis shows how a distributed implementation can be systematically designed from a specification, guided by precise formal goals. While our results are based on formal criteria, we show how violations of each of those criteria can lead to real attacks. We distill the key ideas behind those attacks and propose corrections in terms of useful design principles. We show that other stateful computations can be distributed just as well using those principles. 

%\begin{enumerate}
%\item If the access policy is static, then a secure implementation requires fake capabilities and adequate differentiation of capabilities to prevent capabilities from prematurely leaking information about the access policy. 
%\item If accesses can be dynamically granted but not dynamically revoked, then an implementation following (1) is safe but not secure. (Such an implementation leaks information about the access policy.) 
%\item If accesses can be dynamically revoked, then even a safe implementation requires expiry of capabilities (and a clock). Further:
%\begin{enumerate}
%\item If expiry of capabilities and failure of capabilities are unobservable and accesses cannot be dynamically granted, then a secure implementation requires that the clock also be unobservable. 
%\item Otherwise (in the general case), a secure implementation requires time bounds on processing requests in the specification.
%\end{enumerate}
%\end{enumerate}
%Our full abstraction proofs rely on modular safety proofs; the security definitions and proof techniques that we develop may be of independent interest.
\end{abstract}

%\tableofcontents

%\vspace{-2mm}
\section{Introduction}
\noindent
In most file systems, protection relies on access control. Usually the access checks are local---the file system maintains an access policy that specifies which principals may access which files, and any access to a file is guarded by a local check that enforces the policy for that file. In recent file systems, however, the access checks are distributed, and access control is implemented via cryptographic techniques. 
%Reasoning about the security of such systems becomes much simpler if these implementations preserve the essence of local access control. 
In this paper, we %simply reasoning about the security of such systems by showing 
try to understand the extent to which these distributed implementations of access control preserve the simple character of local access checks. 
%The aim of this paper is to develop this understanding in a systematic way.
%This distribution is guided by one of two roughly complementary design trends. 
%While security and performance remain important considerations in both designs, one of these, \emph{network-attached storage}, focuses on performance under massive request loads; the other, \emph{untrusted storage}, focuses on security under minimal trust assumptions. 

We focus on implementations that appear in file systems based on networked storage~\cite{gobioff-security}. In such systems, access control and storage are parallelized to improve performance. Execution requests are served by storage servers; such requests are guided by access requests that are served elsewhere by access-control servers. %Several operation requests may be guided by a single access request. 
When a user requests access to a file, an access-control server certifies the access decision for that file by providing the user with an unforgeable \emph{capability}. Any subsequent execution request carries that capability as proof of access; a storage server can efficiently verify that the capability is authentic and serve the execution request.

We formally study the correctness of these implementations vis-\`a-vis a simple specification of local access control. % for networked storage. 
Implementing static access policies already requires some care in this setting; dynamic access policies cause further problems that require considerable analysis. We study these cases separately in Sections \ref{intro-stat} and \ref{intro-dyn}. Based on our analysis, we develop formal models and proofs for an implementation of arbitrary access policies in Section \ref{sec:mpdyn}. %{\bf Clarify formal and semi-formal}

%%Violations can be readily exploited to obtain attacks. 
%We show that a secure implementation of a static access policy already requires some care in order to prevent capabilities from prematurely leaking information about the access policy. 
%A dynamic access policy causes further problems. For instance, if accesses can be dynamically granted then such an implementation does not remain secure---it leaks information about the access policy. If accesses can be dynamically revoked then such an implementation does not even remain safe. A safe implementation is possible only if there are time bounds after which capabilities ``expire". A secure implementation is possible only if there are time bounds with similar effects in the specification.

We consider both safety and security for correctness; loosely, safety requires that an implementation does not introduce unspecified behaviors, and security requires that an implementation preserves the specified behavioral equivalences. 
Our proofs of safety and security are built modularly by showing simulations; we develop the necessary definitions and proof techniques in Section \ref{theory}.

Our analysis shows how a distributed implementation can be systematically designed from a specification, guided by precise formal goals. %While our results are based on formal criteria, 
We justify those goals by showing how their violations can lead to real attacks  (Sections \ref{intro-stat} and~\ref{intro-dyn}). Further, we distill the key ideas behind those attacks and propose corrections in terms of useful design principles. %We elaborate on these principles in Section \ref{discuss}. 
We show that other stateful computations can be distributed just as well using those principles (Section \ref{apply}).

%Surprisingly, this setting has received little formal attention. Techniques and proofs in this setting may be expected to apply as well to other distributed authorization frameworks, \emph{e.g.}, trust management and proof-carrying authentication.
%, since the nuances of security therein have received far less formal attention than they merit. 

\paragraph{\em Comparison with related work} This paper culminates a line of work that we begin in~\cite{ChaudhuriAbadi-FMSE05} and continue in ~\cite{ForteChaudhuriA06}. In~\cite{ChaudhuriAbadi-FMSE05}, we show how to securely implement static access policies with capabilities; in~\cite{ForteChaudhuriA06}, we present a safe (but not secure) implementation of dynamic access policies in that setting. In this paper, we carefully review those results, and systematically analyze the difficulties that arise for security in the case of dynamic access policies. Our analysis leads us to develop variants of the implementation in~\cite{ForteChaudhuriA06} that we can prove secure with appropriate assumptions. The proofs are built by a new, instructive technique, which may be of independent interest. % than in \cite{ChaudhuriAbadi-FMSE05,ForteChaudhuriA06}. % that vastly improve their presentation. %so that they are more elegant than the ones in \cite{ChaudhuriAbadi-FMSE05,ForteChaudhuriA06}. 

Further, guided by our analysis of access control, we show how to automatically derive secure distributed implementations of other stateful computations. This approach is reminiscent of secure program partitioning \cite{spp}. % {\bf [Fournet POPL'08], secure program partitioning}%This seems to be the first automatic method for deriving secure distributed protocols from local specifications. 

%This result is not merely a coincidence, and we consider it an important contribution of this paper.
%In doing so, we vastly improve our understanding of the subtleties 

Access control for networked storage has been studied in lesser detail by Gobioff~\cite{gobioff-security} using belief logics, and by Halevi \emph{et al.}~\cite{Halevi05} using universal composability~\cite{ucframework}. The techniques used in this paper are similar to those used by Abadi \emph{et al.} for secure implementation of channel abstractions \cite{abadi98secure} and authentication primitives \cite{authprim}, and by Maffeis to study the equivalence of communication patterns in distributed query systems \cite{Maffeis05}. These techniques rely on  programming languages concepts, including testing equivalence~\cite{nicohenn84} and full abstraction \cite{milnerfullabs,pplt}. 
A huge body of such techniques have been developed for formal specification and verification of systems. 

%Specifically, some other techniques that have been applied to study access control in networked storage include belief logics, used by Gobioff in~\cite{gobioff-security}, and universal composability~\cite{REFuc}, used by Halevi \emph{et al.} in~\cite{Halevi05}. 

We do not consider access control for untrusted storage~\cite{plutus} in this paper. In file systems based on untrusted storage, files are cryptographically secured before storage, and their access keys are managed and shared by users. %Storage servers are not assumed to preserve secrecy or integrity of files. 
%We do not deal with such implementations in this paper---
As such, untrusted storage is quite similar to public communication, and standard techniques for secure communication on public networks apply for secure storage in this setting. 
Related work in that area includes formal analysis of protocols for secure file sharing on untrusted storage \cite{secfsbyz,Blanchet-Chaudhuri}, as well as correctness proofs for the cryptographic techniques  involved in such protocols~\cite{lazyrevcfs,keyregress,keyupd-lazyrev}. 

%\paragraph{Organization of the paper} In Section \ref{intro-stat}, we introduce our correctness criteria and systematically reconstruct the implementation of static access policies in \cite{ChaudhuriAbadi-FMSE05}. In Section \ref{intro-dyn}, we analyze the more difficult problem of implementing dynamic access policies, contrasting the 

\section{Review: the case of static access policies}\label{intro-stat}
\noindent
%{\bf Organization? We use testing equivalence} 
To warm up, let us focus on implementing access policies that are \emph{static}. In this case, a secure implementation already appears in~\cite{ChaudhuriAbadi-FMSE05}. Below we systematically reconstruct that implementation, focusing on a detailed analysis of its correctness. This analysis allows us to distill some basic design principles, marked with bold {\bf R}, in preparation for later sections, where we consider the more difficult problem of implementing dynamic access policies.
% to Section~\ref{intro-dyn}. 
%, assuming that accesses cannot be dynamically granted or revoked.

Consider the following protocol, ${\it NS}^s$, for networked storage.\footnote{By convention, we use superscripts ${}^s$ and ${}^d$ to denote ``static" and ``dynamic", and superscripts ${}^+$ and ${}^-$ to denote ``extension" and ``restriction".} Principals include users $U,V,W\dots$, an access-control server $A$, and a storage server $S$. We assume that $A$ maintains a (static) access policy $F$ and $S$ maintains a store $\rho$. Access decisions under $F$ follow the relation $F \vdash_U \op$ over users $U$ and operations $\op$. Execution of an operation $\op$ under $\rho$ follows the relation $\rho\llbracket\op\rrbracket \Downarrow \rho'\llbracket r\rrbracket$ over next stores $\rho'$ and results $r$. Let $K_{AS}$ be a secret key shared by $A$ and $S$, and $\func{mac}$ be a function over messages and keys that produces unforgeable message authentication codes (MACs) \cite{lecnotes}. We assume that MACs can be decoded to retrieve their messages. (Usually MACs are explicitly paired with their messages, so that the decoding is trivial.)
\[
\left.
\begin{array}{crclcl}
 (1) & U & \rightarrow  & A & : & \op \\
 (2) & A & \rightarrow  & U & : & \func{mac}(\op,K_{AS})~~\mbox{ if }F \vdash_U \op \\
 (2') & A & \rightarrow  & U & : & \mathtt{error}~~\mbox{ otherwise } \\
 \\
% \end{array}
%\right.
%\]
%\[
%\left.
%\begin{array}{crclcl}
 (3) & V & \rightarrow  & S & : & \kappa \\
 (4) & S & \rightarrow  & V & : & r~~\mbox{ if }\kappa = \func{mac}(\op,K_{AS})\\
 &&&&&~~~~\mbox{ and }\rho\llbracket \op\rrbracket \Downarrow \rho'\llbracket r\rrbracket \\
 (4') & S & \rightarrow  & V & : & \mathtt{error}~~\mbox{ otherwise }
 \end{array}
\right.
\]
Here a user $U$ requests $A$ for access to an operation $\op$, and $A$ returns a capability for $\op$ only if $F$ specifies that $U$ may access $\op$. Elsewhere, a user $V$ requests $S$ to execute an operation by sending a capability $\kappa$, and $S$ executes the operation only if $\kappa$ authorizes access to that operation.

What does ``safety" or ``security" mean in this setting? A reasonable specification of correctness is the following trivial protocol, ${\it IS}^s$, for ideal storage. Here principals include users $U,V,W,\dots$ and a server $D$. The access policy $F$ and the store $\rho$ are both maintained by $D$; the access and execution relations remain as above. There is no cryptography. %We then ask whether ${\it NS}^s$ \emph{implements} ${\it IS}^s$ under standard notions of implementation correctness. 
\[
\left.
\begin{array}{crclcl}
({\rm i}) &  V & \rightarrow  & D & : & \op \\
({\rm ii}) &  D & \rightarrow  & V & : & r~~\mbox{ if }F \vdash_V \op\mbox{ and }\rho\llbracket \op\rrbracket \Downarrow \rho' \llbracket r\rrbracket \\
({\rm ii'}) &  D & \rightarrow  & V & : & \mathtt{error}~~\mbox{ otherwise }
\end{array}
\right.
\]
Here a user $V$ requests $D$ to execute an operation $\op$, and $V$ executes $\op$ only if $F$ specifies that $V$ may access $\op$. This trivial protocol is correct ``by definition"; so if ${\it NS}^s$ implements this protocol, it is correct as well. 

What notions of implementation correctness are appropriate here? A basic criterion is that of safety \cite{refmap}. 
\begin{definition}[Safety]
Under any context (adversary), the behaviors of a safe implementation are included in the behaviors of the specification.
\end{definition}
\noindent
In practice, a suitable notion of inclusion may need to be crafted to accommodate specific implementation behaviors by design (such as those due to messages $(1)$, $(2)$, and $(2')$ in ${\it NS}^s$). Typically, those behaviors can be eliminated by a specific context (called a ``wrapper"), and safety may be defined modulo that context as long as other, interesting behaviors are not eliminated. 

Still, safety only implies the preservation of certain trace properties. A more powerful criterion may be derived from the programming languages concept of semantics preservation, otherwise known as \emph{full abstraction} \cite{milnerfullabs,pplt}. 
%\vspace{-0.5mm}
\begin{definition}[Security]
A secure implementation preserves behavioral equivalences of the specification. 
\end{definition}
\noindent
In this paper, we tie security to an appropriate may testing congruence~\cite{nicohenn84}.  %\footnote{A congruence is an equivalence that is closed under arbitrary contexts.}
We consider a protocol instance to include the file system and some code run by ``honest" users, and assume that an arbitrary context colludes with the remaining ``dishonest" users. From any ${\it NS}^s$ instance, we derive its ${\it IS}^s$ instance by an appropriate refinement map~\cite{refmap}. % (\emph{e.g.}, one induced by may testing); 
If ${\it NS}^s$ securely implements ${\it IS}^s$, then for all ${\it NS}^s$ instances $Q_1$ and $Q_2$,  $Q_1$ and $Q_2$ are congruent if their ${\it IS}^s$ instances are congruent. 
%Security of ${\it NS}^s$ then reduces to the following question:
%%
%\vspace{-0.5mm}
%\begin{quote}{\em Is ${\it NS}^s$ a secure implementation of ${\it IS}^s$?}
%\vspace{-1mm}
%\end{quote}
%

Security implies safety for all practical purposes, so a safety counterexample usually suffices to break security. For instance, we are in trouble if operations that cannot be executed in ${\it IS}^s$ can somehow be executed in ${\it NS}^s$ by manipulating capabilities. Suppose that $F \not\vdash_V \op$ for all dishonest $V$. Then no such $V$ can execute $\op$ in ${\it IS}^s$. Now suppose that some such $V$ requests execution of $\op$ in ${\it NS}^s$. We know that $\op$ is executed only if $V$ shows a capability $\kappa$ for $\op$. Since $\kappa$ cannot be forged, it must be obtained from $A$ by some honest $U$ that satisfies $F \vdash_U \op$. %On the other hand, the adversary \emph{can} execute $\op$ in ${\it NS}^s$ if some honest $U$ that satisfies $F \vdash_U \op$ obtains a capability $\kappa$ for $\op$, and passes $\kappa$ to $V$. 
Therefore:
%\vspace{-0.5mm}
\begin{description}
\item[\R]\label{Rsecret} Capabilities obtained by honest users must not be shared with dishonest users.
\end{description}
(However $U$ can still share $\kappa$ with honest users, and any execution request with $\kappa$ can then be reproduced in the specification as an execution request by $U$.)

While (R\ref{Rsecret}) prevents \emph{explicit} leaking of capabilities, we in fact require that capabilities do not leak \emph{any} information that is not available to ${\it IS}^s$ contexts. Information may also be leaked implicitly (by observable effects). Therefore:
%\vspace{-0.5mm}
\begin{description}
\item[\R]\label{Rexamine} Capabilities obtained by honest users must not be examined or compared. % before their use. 
\end{description}
Both (R\ref{Rsecret}) and (R\ref{Rexamine}) may be enforced by typechecking the code run by honest users. 

Finally, we require that information is not leaked via capabilities obtained by dishonest users. (Recall that such capabilities are already available to the adversary.) Unfortunately, a capability for an operation $\op$ is provided \emph{only} to those users who have access to $\op$ under $F$; in other words, $A$ leaks information on $F$ whenever it returns a capability! %\footnote{If we do not care, we must allow the same leak in the specification.} % See Section \ref{discuss}.} 
This leak breaks security.
Why? Consider implementation instances $Q_1$ and $Q_2$ with $\op$ as the only operation, whose execution returns $\mathtt{error}$ and may be observed only by honest users; suppose that a dishonest user has access to $\op$ in $Q_1$ but not in $Q_2$. Then $Q_1$ and $Q_2$ can be distinguished by a context that requests a capability for $\op$---a capability will be returned in $Q_1$ but not in $Q_2$---but their specification instances cannot be distinguished by any context. 

Why does this leak concern us? After all, we expect that executing an operation \emph{should} eventually leak some information about access to that operation, since otherwise, having access to that operation is useless! However the leak here is premature; it allows a dishonest user to obtain information about its access to $\op$ in an undetectable way, \emph{without} having to request execution of $\op$.

To prevent this leak, we must modify the protocol:
\begin{description}
\item[\R]\label{Rfake} ``Fake" capabilities for $\op$ must be returned to users who do not have access to $\op$. 
\end{description}
The point is that it should not be possible to distinguish the fake capabilities from the real ones prematurely. Let $\overline K_{AS}$ be another secret key shared by $A$ and $S$. As a preliminary fix, let us modify the following message in ${\it NS}^s$. 
%\vspace{-1mm}
\[
\left.
\begin{array}{crclcl}
 (2') & A & \rightarrow  & U & : & \func{mac}(\op,\overline K_{AS})~~\mbox{ if }F \not\vdash_U \op 
\end{array}
\right.
\]
Unfortunately this modification is not enough, since the adversary can still compare capabilities that are obtained by different users for a particular operation $\op$, to know if their accesses to $\op$ are the same under $F$. To prevent this leak:%\footnote{We use \~~to mark design choices that are more ``exploratory" than others; alternative choices may exist.}
\begin{description}
\item[\R]\label{Rdiff} Capabilities for different users must be different.
\end{description}
For instance, a capability can mention the user whose access it authenticates.
Making the meaning of a message explicit in its content is a good design principle for security \cite{Abadi96}, and we use it on several occasions in this paper. %(Alternatively, in this case it suffices to include a fresh nonce in every capability.)
%Now capabilities for $\op$ that are obtained by different users differ irrespective of their accesses to $\op$ under $F$. 
Accordingly we modify the following messages in ${\it NS}^s$. \[
\left.
\begin{array}{crclcl}
 (2) & A & \rightarrow  & U & : & \func{mac}(\tup{U,\op},K_{AS})~~\mbox{ if }F \vdash_U \op \\
 (2') & A & \rightarrow  & U & : & \func{mac}(\tup{U,\op},\overline K_{AS})~~\mbox{ otherwise }\\
 \\
 (4) & S & \rightarrow  & V & : & r~~\mbox{ if }\kappa = \func{mac}(\tup{\_,\op},K_{AS}) \\
  &&&&&~~~~\mbox{ and }\rho\llbracket\op\rrbracket \Downarrow \rho'\llbracket r\rrbracket 
\end{array}
\right.
\]
(On receiving a capability $\kappa$ from $V$, $S$ still does not care whether $V$ is the user to which $\kappa$ is issued, even if that information can now be obtained from $\kappa$.)

The following result can then be proved (\emph{cf.} \cite{ChaudhuriAbadi-FMSE05}).
%\vspace{-1mm}
\begin{theorem}\label{thm:static}
${\it NS}^s$ securely implements ${\it IS}^s$.
\end{theorem}

%(see Section \ref{sec:mpdyn}): 

\section{The case of dynamic access policies}\label{intro-dyn}
\noindent
We now consider the more difficult problem of implementing dynamic access  policies. 
Let $F$ be dynamic; the following protocol, ${\it NS}^d$, is obtained by adding administration messages to ${\it NS}^s$. Execution of an administrative operation $\theta$ under $F$ follows the relation $F\llbracket\theta\rrbracket \Downarrow F'\llbracket r\rrbracket$ over next policies $F'$ and results $r$. 
%\vspace{-1mm}
\[
\left.
\begin{array}{crclcl}
\!\!\!\! (5)\! & W & \rightarrow  & A & : & \theta \\
\!\!\!\! (6)\! & A & \rightarrow  & W & : & r~~\mbox{ if }F \vdash_W \theta\mbox{ and }F\llbracket\theta\rrbracket \Downarrow F'\llbracket r\rrbracket \\
\!\!\!\! (6')\! & A & \rightarrow  & W & : & \mathtt{error}~~\mbox{ otherwise }
\end{array}
\right.
\]
Here $A$ executes $\theta$ (perhaps modifying $F$) if $F$ specifies that $W$ controls~$\theta$. 
The following protocol, ${\it IS}^d$, is obtained by adding similar messages to ${\it IS}^s$.
%\vspace{-1mm}
\[
\left.
\begin{array}{crclcll}
 \!\!\!\!({\rm iii})\! & W & \rightarrow  & D & : & \theta \\
 \!\!\!\!({\rm iv})\! & D & \rightarrow  & W & : & r~~\mbox{ if }F \vdash_W \theta\mbox{ and }F\llbracket\theta\rrbracket \Downarrow F'\llbracket r\rrbracket \\
\!\!\!\! ({\rm iv'})\! & D & \rightarrow  & W & : & \mathtt{error}~~\mbox{ otherwise } 
%\vspace{-1.5mm}
\end{array}
\right.
\]
Unfortunately ${\it NS}^d$ does not remain secure with respect to ${\it IS}^d$.
Consider the ${\it NS}^d$ pseudo-code below. Informally, $\mathtt{acquire}~\kappa$ means ``obtain a capability $\kappa$" and $\mathtt{use}~\kappa$ means ``request execution with $\kappa$"; $\mathtt{chmod}~\theta$ means ``request access modification $\theta$"; and $\mathtt{success}$ means ``detect successful use of a capability". 
Here $\kappa$ is a capability for an operation $\op$ and $\theta$ modifies access to $\op$. 
\begin{description}
\item[t1] %\fbox{
$\mathtt{acquire}~\kappa;%$} $
 \mathtt{chmod}~\theta; \mathtt{use}~\kappa; \mathtt{success}$ 
\item[t2] $\mathtt{chmod}~\theta;%$ \fbox{$ 
\mathtt{acquire}~\kappa;%$} $ 
\mathtt{use}~\kappa; \mathtt{success}$
\end{description}
Now (t1) and (t2) map to the same ${\it IS}^d$ pseudo-code
$\mathtt{chmod}~\theta; \mathtt{exec}~\op; \mathtt{success}$---informally, $\mathtt{exec}~\op$ means ``request execution of $\op$". Indeed, requesting execution with $\kappa$ in ${\it NS}^d$ amounts to requesting execution of $\op$ in ${\it IS}^d$, so the refinement map must erase instances of $\mathtt{acquire}$ and replace instances of $\mathtt{use}$ with the appropriate instances of $\mathtt{exec}$.
However, suppose that initially no user has access to $\op$, and $\theta$ specifies that all users may access $\op$. Then (t1) and (t2) can be distinguished by testing the event $\mathtt{success}$. In (t1) $\kappa$ does not authorize access to $\op$, so $\mathtt{success}$ must be false; but in (t2) $\kappa$ may authorize access to $\op$, so $\mathtt{success}$ may be true. 

Moreover, if revocation is possible, ${\it NS}^d$ does not even remain safe with respect to ${\it IS}^d$! Why? Let $\theta$ specify that access to $\op$ is revoked for some user $U$, and $\mathtt{revoked}$ be the event that $\theta$ is executed (thus modifying the access policy). In ${\it IS}^d$, $U$ cannot execute $\op$ after $\mathtt{revoked}$. But in ${\it NS}^d$, $U$ can execute $\op$ after $\mathtt{revoked}$ by using a capability that it acquires before $\mathtt{revoked}$.  

%\vspace{-3mm}
\paragraph{\em Safety in a special case}
One way of eliminating the counterexample above is to make the following assumption:
\begin{description}
\item[\A]\label{Arevoke} Accesses cannot be dynamically revoked.
\end{description}
We can then prove the following new result (see Section \ref{sec:mpdyn}).
%\vspace{-1mm}
\begin{theorem}
${\it NS}^d$ safely implements ${\it IS}^d$ assuming {\rm (A\ref{Arevoke})}.\footnote{Some implementation details, such as (R\ref{Rfake}), are not required for safety.} 
\end{theorem}
\noindent
The key observation is that with (A\ref{Arevoke}), a user $U$ cannot access $\op$ until it can always access $\op$, so $U$ gains no advantage by acquiring capabilities early.

%But without (A\ref{Arevoke}), 

%\vspace{-3mm}
\paragraph{\em Safety in the general case}
%Clearly we need a new idea in the general case.
Safety breaks with revocation. However, we can recover
safety by introducing \emph{time}. % and having capabilities carry time bounds after which they expire. 
Let $A$ and $S$ share a logical clock (or counter) that measures time, and let the same clock appear in $D$. We have that:
\begin{description}
\item[\R]\label{Rexpire} Any capability that is produced at time $\clk$ expires at time $\clk+1$.%\footnote{This interval can be adjusted, as discussed in Section \ref{discuss}.} 
%\vspace{-1mm}
\item[\R]\label{Rdelay} Any administrative operation requested at time $\clk$ is executed at the next clock tick (to time $\clk + 1$), so that policies in ${\it NS}^d$ and ${\it IS}^d$ may change only at clock ticks (and not between).  
\end{description}

\noindent
We call this arrangement a ``midnight-shift scheme", since the underlying idea is the same as that of periodically shifting guards at a museum or a bank. Implementing this scheme is straightforward. For (R\ref{Rexpire}), capabilities carry timestamps. For (R\ref{Rdelay}), administrative operations are executed on an ``accumulator" $\Xi$ instead of $F$, and at every clock tick, $F$ is updated to $\Xi$. Accordingly, we modify the following messages in ${\it NS}^d$ to obtain the protocol ${\it NS}^{d+}$. 
\[
\left.
\begin{array}{crclcl}
\!\!\!\! (2)\!\! & A & \rightarrow  & U\!\!\! & \!\!:\!\! & \!\func{mac}(\tup{U,\op,\clk},K_{AS})~~\mbox{ if }F \vdash_U \op \\
\!\!\!\! (2')\!\! & A & \rightarrow  & U\!\!\! & \!\!:\!\! & \!\func{mac}(\tup{U,\op,\clk},\overline K_{AS})~~\mbox{ otherwise }\\
\\
\!\!\!\! (4)\!\! & S & \rightarrow  & V\!\!\! & \!\!:\!\! & \!r~~\mbox{ if }\kappa = \func{mac}(\tup{\_,\op,\clk},K_{AS}) \\
 &&&&&~~~~\mbox{ and }\rho\llbracket\op\rrbracket \Downarrow \rho'\llbracket r\rrbracket \\
 \\
\!\!\!\! (6)\!\! & A & \rightarrow  & W\!\!\! & \!\!:\!\! & \!r~~\mbox{ if }F\vdash_W \theta\mbox{ and }\Xi\llbracket\theta \rrbracket \Downarrow \Xi'\llbracket r\rrbracket
\end{array}
\right.
\]
Likewise, we modify the following message in ${\it IS}^d$ to obtain the protocol ${\it IS}^{d+}$.
 \[
\left.
\begin{array}{crclcl}
  \!\!\!\!({\rm iv})\! & D & \rightarrow  & W & \!\!:\!\! & r~~\mbox{ if }F\vdash_W \theta\mbox{ and }\Xi\llbracket\theta \rrbracket \Downarrow \Xi'\llbracket r\rrbracket 
\end{array}
\right.
\]
Now a capability that carries $\clk$ as its timestamp certifies a particular access decision at the instant $\clk$: the meaning is made explicit in the content, which is good practice. However, recall that MACs can be decoded to retrieve their messages. In particular, one can tell the time in ${\it NS}^{d+}$ by decoding capabilities. Clearly we require that:
\begin{description}
\item[\R]\label{Rtime} If it is possible to tell the time in ${\it NS}^{d+}$, it must also be possible to do so in ${\it IS}^{d+}$.
\end{description}
So we must make it possible to tell the time in ${\it IS}^{d+}$. 
(The alternative is to make it impossible to tell the time in ${\it NS}^{d+}$, by encrypting the timestamps carried by capabilities. Recall that the notion of ``time" here is purely logical.) 
Accordingly we add the following messages to ${\it IS}^{d+}$.
\[
\left.
\begin{array}{crclcl}
%  ({\rm iv}) & D & \rightarrow  & W & : & r & \mbox{if }F\vdash_W \op\mbox{ and }\Xi\llbracket\op \rrbracket \Downarrow \Xi'\llbracket r\rrbracket \\
%\\
 ({\rm v}) & U & \rightarrow  & D & : & () \\
 ({\rm vi}) & D & \rightarrow  & U & : & \clk 
\end{array}
\right.
\]
The following result can then be proved (\emph{cf.} \cite{ForteChaudhuriA06}). %(Section \ref{sec:mpdyn}):
%\vspace{-1mm}
\begin{theorem}
${\it NS}^{d+}$ safely implements ${\it IS}^{d+}$.
\end{theorem}
\noindent
This result appears in~\cite{ForteChaudhuriA06}. Unfortunately, the definition of safety in~\cite{ForteChaudhuriA06} is rather non-standard. Moreover, beyond this result, security is not considered in \cite{ForteChaudhuriA06}. In the rest of this section, we analyze the difficulties that arise for security, and present new results. 

%\vspace{-5mm}
%\paragraph{But what about security?}
It turns out that there are several recipes to break security, and expiry of capabilities is a common ingredient. Clearly, using an expired capability has no counterpart in ${\it IS}^{d+}$. So:
%\vspace{-1mm}
\begin{description}
\item[\R]\label{Rstale} Any use of an expired capability must block (without any observable effect). 
\end{description}
Indeed, security breaks without (R\ref{Rstale}). Consider the ${\it NS}^{d+}$ pseudo-code below. Informally, $\mathtt{stale}$ means ``detect any use of an expired capability". Here $\kappa$ is a capability for operation $\op$.
%\vspace{-5mm}
\begin{description}
\item[t3] $\mathtt{acquire}~\kappa; \mathtt{use}~\kappa; \mathtt{stale}$
\end{description}
\noindent
Without (R\ref{Rstale}), (t3) can be distinguished from a $\mathtt{false}$ event by testing the event $\mathtt{stale}$. But consider implementation instances $Q_1$ and $Q_2$ with $\op$ as the only operation, whose execution has no observable effect on the store; let $Q_1$ run (t3) and $Q_2$ run $\mathtt{false}$. Since $\mathtt{stale}$ cannot be reproduced in the specification, it must map to $\mathtt{false}$. So the specification instances of $Q_1$ and $Q_2$ run $\mathtt{exec}~\op; \mathtt{false}$ and $\mathtt{false}$. These instances cannot be distinguished. %Thus security breaks without (R\ref{Rstale}). 

Moreover, expiry of a capability yields the information that time has elapsed between the acquisition and use of that capability. We may expect that leaking this information is harmless; after all, the elapse of time can be trivially detected by inspecting timestamps. Then why should we care about such a leak? If the adversary knows that the clock has ticked at least once, it also knows that any pending administrative operations have been executed, possibly modifying the access policy. If this information is leaked in a way that cannot be reproduced in the specification, we are in trouble. %! Of course,; we are only concerned about non-trivial ways of obtaining this information, which cannot be reproduced un the specification.
Any such way allows the adversary to \emph{implicitly} control the expiry of a capability before its use. (Explicit controls, such as comparison of timestamps, are not problematic, since they can be reproduced in the specification.) 

For instance, consider the ${\it NS}^{d+}$ pseudo-code below. 
Here $\kappa$ and $\kappa'$ are capabilities for operations $\op$ and $\op'$, and $\theta$ modifies access to $\op$. 
%\vspace{-1mm}
\begin{description}
\item[t4] %\fbox{
$\mathtt{acquire}~\kappa';$

$\!\!\!\!\mathtt{chmod}~\theta; \mathtt{acquire}~\kappa; \mathtt{use}~\kappa; \mathtt{success};$

$\!\!\!\!\mathtt{use}~\kappa'; \mathtt{success}$
%\vspace{-1mm}
\item[t5] $\mathtt{chmod}~\theta; \mathtt{acquire}~\kappa; \mathtt{use}~\kappa; \mathtt{success};$

%\fbox{
$\!\!\!\!\mathtt{acquire}~\kappa';%$} $
  \mathtt{use}~\kappa'; \mathtt{success}$
\end{description}
Both (t4) and (t5) map to the same ${\it IS}^{d+}$ pseudo-code $$\mathtt{chmod}~\theta; \mathtt{exec}~\op; \mathtt{success}; \mathtt{exec}~\op'; \mathtt{success}$$ 
But suppose that initially no user has access to $\op$ and all users have access to $\op'$, and $\theta$ specifies that all users may access $\op$. The intermediate $\mathtt{success}$ event is true only if $\theta$ is executed; therefore it ``forces" time to elapse for progress. Now (t4) and (t5) can be distinguished by testing the final $\mathtt{success}$ event. In (t4) $\kappa'$ must be stale when used, so the event must be false; but in (t5) $\kappa'$ may be fresh when used, so the event may be true. Therefore, security breaks.

%We return to this counterexample on several occasions below. Let us now look at some possible solutions.

%\vspace{-3mm}
\paragraph{\em Security in a special case}
One way of plugging such leaks is to consider that the elapse of time is altogether unobservable. (This prospect is not as shocking as it sounds, since here ``time" is simply the value of a privately maintained counter.) 

We expect that executing an operation has some observable effect. Now if initially a user does not have access to an operation, but that access can be dynamically granted, then the elapse of time \emph{can} be detected by observing the effect of executing that operation. So we must assume that:
%\vspace{-1mm}
\begin{description}
\item[\A]\label{Agrant} Accesses cannot be dynamically granted.
\end{description}
On the other hand, we must allow accesses to be dynamically revoked, since otherwise the access policy becomes static. Now if initially a user has access to an operation, but that access can be dynamically revoked, then it is possible to detect the elapse of time if the \emph{failure} to execute that operation is observable. So we must assume that:
%, Moreover, while execution of operations must have observable effects, we may consider an implementation where conversely:
%\vspace{-1mm}
\begin{description}
\item[\A]\label{Ablock} Any unsuccessful use of a capability blocks (without any observable effect).
\end{description}
Let us now try to adapt the counterexample above with (A\ref{Agrant}) and (A\ref{Ablock}). Suppose that initially all users have access to $\op$ and $\op'$, and $\theta$ specifies that no user may access $\op$. Consider the ${\it NS}^{d+}$ pseudo-code below. Informally, $\mathtt{failure}$ means ``detect unsuccessful use of a capability". 
%\vspace{-1mm}
\begin{description}
\item[t6] %\fbox{
$\mathtt{acquire}~\kappa';$

$\!\!\!\!\mathtt{chmod}~\theta; \mathtt{acquire}~\kappa; \mathtt{use}~\kappa; \mathtt{failure};$

$\!\!\!\!\mathtt{use}~\kappa'; \mathtt{success}$
%\vspace{-1mm}
\item[t7] $\mathtt{chmod}~\theta; \mathtt{acquire}~\kappa; \mathtt{use}~\kappa; \mathtt{failure}; $

%\fbox{
$\!\!\!\!\mathtt{acquire}~\kappa';%$} $
 \mathtt{use}~\kappa'; \mathtt{success}$
\end{description}
Both (t6) and (t7) map to the same ${\it IS}^{d+}$ pseudo-code $$\mathtt{chmod}~\theta; \mathtt{exec}~\op; \mathtt{failure}; \mathtt{exec}~\op'; \mathtt{success}$$
Fortunately, now (t6) and (t7) cannot be distinguished, since the intermediate $\mathtt{failure}$ event cannot be observed if true. (In contrast, recall that the intermediate $\mathtt{success}$ event in (t4) and (t5) forces a distinction between them.)

Indeed, with (A\ref{Agrant}) and (A\ref{Ablock}) there remains no way to detect the elapse of time, except by comparing timestamps. To prevent the latter, we
 assume that:
%\vspace{-1mm}
\begin{description}
\item[\A]\label{Aclock} Timestamps are encrypted.%\footnote{Hiding the clock is futile if the clock is not privately synchronized.}
\end{description}
Let $E_{AS}$ be a secret key shared by $A$ and $S$. The encryption of a term $M$ with $E_{AS}$ under a random coin $m$ is written as $\{m,M\}_{E_{AS}}$. %, and also allows fake capabilities to be simplified. 
We remove  message $(4')$ and modify the following messages in ${\it NS}^{d+}$ to obtain the protocol ${\it NS}^{d-}$. (Note that randomization takes care of (R\ref{Rdiff}), so capabilities are not \emph{required} to mention users here.) 
\[
\left.
\begin{array}{crclcl}
\!\!\!\! (2)\!\! &\! A & \rightarrow  & U \!\!\!& \!\!\!:\!\!\! &\! \func{mac}(\tup{U,\op,\{m,\clk\}_{E_{AS}}},K_{AS})\\
&&&&&~~~~~~~~~~~~~~~~~~~~~~~~~~~~~~~~~~~~~~~~~\mbox{ if }F \vdash_U \op \\
\!\!\!\! (2')\!\! &\! A & \rightarrow  & U \!\!\!& \!\!\!:\!\!\! &\! \func{mac}(\tup{U,\op,\{m,\clk\}_{E_{AS}}},\overline K_{AS})\\
&&&&&~~~~~~~~~~~~~~~~~~~~~~~~~~~~~~~~~~~~~~~~~\mbox{ otherwise} \\
\\
\!\!\!\! (4)\!\! &\! S & \rightarrow  & V \!\!\!& \!\!\!:\!\!\! &\! r~~\mbox{ if }\kappa = \func{mac}(\tup{\_,\op,\mathcal T},K_{AS})\mbox{,} \\
 &&&&&~~~~\mathcal T = \{\_,\clk\}_{E_{AS}}\mbox{, and }\rho\llbracket\op\rrbracket \Downarrow \rho'\llbracket r\rrbracket 
\end{array}
\right.
\]
%(R\ref{Rtime}) is already taken care of; 
Accordingly, we remove the messages $({\rm iv'})$, $({\rm v})$, and $({\rm vi})$ from ${\it IS}^{d+}$ to obtain the protocol ${\it IS}^{d-}$. We can then prove the following new result (see Section \ref{sec:mpdyn}):
\begin{theorem} ${\it NS}^{d-}$ securely implements ${\it IS}^{d-}$ assuming {\rm (A\ref{Agrant})}, {\rm (A\ref{Ablock})}, and {\rm (A\ref{Aclock})}.
\end{theorem}
\noindent
The key observation is that with (A\ref{Agrant}), (A\ref{Ablock}), and (A\ref{Aclock}), time can stand still (so that capabilities never expire). 

%\vspace{-4mm}
\paragraph{\em Security in the general case}
More generally, we may consider plugging problematic leaks by static analysis. (Any such analysis must be incomplete because of the undecidability of the problem.) However, several  complications arise in this case. 
%\vspace{-1mm}
\begin{mylist}
\item The adversary can control the elapse of time by interacting with honest users in subtle ways. Such interactions lead to counterexamples of the same flavor as the one with (t4) and (t5) above, but are difficult to prevent statically without severely restricting the code run by honest users. For instance, even if the suspicious-looking pseudo-code $\mathtt{chmod}~\theta; \mathtt{acquire}~\kappa; \mathtt{use}~\kappa; \mathtt{success}$ in (t4) and (t5) is replaced by an innocuous pair of inputs on a public channel $c$, the adversary can still run the same code in parallel and serialize it by a pair of outputs on $c$ (which serve as ``begin/end" signals).
\item Even if we restrict the code run by honest users, such that every use of a capability can be serialized immediately after its acquisition, % (making it impossible to force time to elapse in between) 
the adversary can still force time to elapse \emph{after} a capability is sent to the file system and \emph{before} it is examined. Unless we have a way to constrain this elapse of time, we are in trouble. 
\end{mylist}
To see how the adversary can break security by interacting with honest users, consider the ${\it NS}^{d+}$ pseudo-code below. Here $\kappa$ is a capability for operation $\op$, and $\theta$ modifies access to $\op$; further $c()$ and $\overline w\tup{}$ denote input and output on public channels $c$ and $w$. 
\begin{description}
\item[t8] $\mathtt{acquire}\:\kappa; \mathtt{use}\:\kappa; c(); \mathtt{chmod}\:\theta; c(); \mathtt{success}; \overline w\tup{}$
\item[t9] $c(); c(); \overline w\tup{}$
\end{description}
Although $\mathtt{use}\:\kappa$ immediately follows $\mathtt{acquire}\:\kappa$ in (t8), the delay between $\mathtt{use}\:\kappa$ and $\mathtt{success}$ can be detected by the adversary to force time to elapse between those events. Suppose that initially no user has access to $\op$ or $\op'$, $\theta$ specifies that a honest user $U$ may access $\op$, and $\theta'$ specifies that all users may access $\op'$. Consider the following context. Here $\kappa'_0$ and $\kappa'_1$ are capabilities for $\op'$.
\begin{description}
\item $\overline c\tup{}; \mathtt{acquire}\:\kappa'_0; \mathtt{use}\:\kappa'_0; \mathtt{failure};$

$\:\mathtt{chmod}\:\theta'; \mathtt{acquire}\:\kappa'_1; \mathtt{use}\:\kappa'_1; \mathtt{success}; \overline c\tup{}$
\end{description}
This context forces time to elapse between a pair of outputs on $c$. The context can distinguish (t8) and (t9) by testing output on $w$: in (t8) $\kappa$ does not authorize access to $\op$, so $\mathtt{success}$ is false and there is no output on $w$; on the other hand, in (t9) there is. Security breaks as a consequence. Consider implementation instances $Q_1$ and $Q_2$ with $U$ as the only honest user and $\op$ and $\op'$ as the only operations, such that only $U$ can detect execution of $\op$ and all users can detect execution of $\op'$; %, while all users can detect execution of $\op'$; 
let $Q_1$ run (t8) and $Q_2$ run (t9). The specification instances of $Q_1$ and $Q_2$ run $\mathtt{exec}\:\op; c(); \mathtt{chmod}\:\theta; c(); \mathtt{success}; \overline w\tup{}$ and $c(); c(); \overline w\tup{}$, which cannot be distinguished: the execution of $\op$ can always be delayed until $\theta$ is executed, so that $\mathtt{success}$ is true and there is an output on $w$. Intuitively, an execution request in ${\it NS}^{d+}$ commits to a time bound (specified by the timestamp of the capability used for the request) within which that request must be processed for progress; but operation requests in ${\it IS}^{d+}$ make no such commitment. 

%We do not assume that noninterference (under some notion of purity or atomicity) between acquisition, use, and validation of capabilities.
%Since (A\ref{Aclock}) no longer helps, we remove that assumption. 
To solve this problem, we must assume that:
\begin{description}
\item[\A]\label{Abound} In ${\it IS}^{d+}$ a time bound is specified for every operation request, so that the request is dropped if it is not processed within that time bound. 
\end{description}
Usual (unrestricted) requests now carry a time bound $\infty$. Accordingly we modify the following messages in ${\it IS}^{d+}$.
\[
\left.
\begin{array}{crclcl}
({\rm i}) &  V & \rightarrow  & D & : & (\op,T) \\
({\rm ii}) &  D & \rightarrow  & V & : & r~~\mbox{ if }\clk \leq T\mbox{, }\\
&&&&&~~~~~F \vdash_V \op\mbox{, and }\rho\llbracket\op\rrbracket \Downarrow \rho'\llbracket r\rrbracket 
\end{array}
\right.
\]
With (A\ref{Abound}), using an expired capability now has a counterpart in ${\it IS}^{d+}$. Informally, if a capability for an operation $\op$ is produced at time $T$ in ${\it NS}^{d+}$, then any use of that capability in ${\it NS}^{d+}$ maps to an execution request for $\op$ in ${\it IS}^{d+}$ with time bound $T$. There remains no fundamental difference between ${\it NS}^{d+}$ and ${\it IS}^{d+}$. 
We can then prove our main new result (see Section \ref{sec:mpdyn}):
\begin{theorem}[Main theorem] ${\it NS}^{d+}$ securely implements ${\it IS}^{d+}$ assuming {\rm (A\ref{Abound})}.\footnote{This result holds with or without (R\ref{Rstale}).}
\end{theorem}
\noindent 
Fortunately, (A\ref{Abound}) seems to be a reasonable requirement, and we impose that requirement implicitly in the sequel. 

\paragraph{\em Discussion}
Let us now revisit the principles developed in Sections \ref{intro-stat} and \ref{intro-dyn}, and discuss some alternatives.

First recall (R\ref{Rfake}), where we introduce fake capabilities to prevent premature leaks of information about the access policy $F$. It is reasonable to consider that we do not care about such leaks, and wish to keep the original message  $(2')$ in ${\it NS}^s$. But then we must allow those leaks in the specification. For instance, we can make $F$ public. 
%More practically, we can add the following messages to ${\it IS}^s$.
% \[
%\left.
%\begin{array}{crclcll}
%({\rm vii}) &  C & \rightarrow  & S & : & \op \\
%({\rm viii}) &  D & \rightarrow  & U & : & L & L = \mathtt{true}\mbox{ if }F \vdash_U \op\mbox{, }=\mathtt{false}\mbox{ otherwise}
%\end{array}
%\right.
%\]
%A user $U$ obtains relevant access information on $\op$ by ``opening" $\op$ in ${\it IS}^s$.
%Consequently, we can modify the following message in ${\it NS}^s$.
% \[
%\left.
%\begin{array}{crclcll}
%({\rm 2'}) &  A & \rightarrow  & U & : & \mathtt{error} & \mbox{if }F \not\vdash_U \op 
%\end{array}
%\right.
%\]
More practically, we can add messages to ${\it IS}^s$ that allow a user to know whether it has access to a particular operation.
% fake capabilities, problems due to expiry of capabilities remain; for instance, (A\ref{Abound}) is still required for security in the case of dynamic access policies.
%
%\begin{description}
%\item[R3] $M$ must return a \emph{fake} capability for $\op$ to users who do not have access to $\op$ under $F$. 
%\end{description}

Next recall (R\ref{Rexpire}) and (R\ref{Rdelay}), where we introduce the midnight-shift scheme. This scheme can be relaxed to allow different capabilities to expire after different intervals, so long as administrative operations that affect their correctness are not executed before those intervals elapse. Let $\func{delay}$ be a function over users $U$, operations $\op$, and clock values $\clk$ that produces time intervals. We may have that:
\begin{description}
\item[R\ref{Rexpire}] Any capability for $U$ and $\op$ that is produced at time $\clk$ expires at time $\clk+\func{delay}(U,\op,\clk)$.
%\vspace{-1mm}
\item[R\ref{Rdelay}] If an administrative operation affects the access decision for $U$ and $\op$ and is requested in the interval $\clk,\dots,\clk+\func{delay}(U,\op,\clk) - 1$, it is executed at the clock tick to time $\clk + \func{delay}(U,\op,\clk)$.  
\end{description}
This scheme remains sound, since any capability for $U$ and $\op$ that is produced at $\clk$ and expires at $\clk + \func{delay}(U,\op,\clk)$ certifies a correct access decision for $U$ and $\op$ between $\clk,\dots,\clk+\func{delay}(U,\op,\clk) - 1$.

Finally, the implementation details in Sections~\ref{intro-stat}~and~\ref{intro-dyn} are far from unique. Guided by the same underlying principles, we can design capabilities in various other ways. For instance, we may have an implementation that does not require $\overline K_{AS}$: any capability is of the form $\func{mac}(\tup{\tup{U,\op,\clk},\{m,L\}_{E_{AS}}},K_{AS})$, where $m$ is a fresh nonce and $L$ is the predicate $F \vdash_U \op$.
%\vspace{-2mm}
%\[
%L = \left\{
%\begin{array}{ll}
%\func{true} & \mbox{if }F \vdash_U \op \\
%\func{false} & \mbox{otherwise }
%\end{array}
%\right.
%\]
Although this design involves more cryptography than the one in ${\it NS}^{d+}$, it reflects better practice: the access decision for $U$ and $\op$ under $F$ is explicit in the content of any capability that certifies that decision. %(The decision is encrypted, for obvious reasons.) 
What does this design buy us? Consider applications where the access decision is not a boolean, but a label, a decision tree, or some arbitrary data structure. 
The design in ${\it NS}^{d+}$ requires a different signing key for each value of the access decision. Since the number of such keys may be infinite, verification of capabilities becomes very inefficient. The design above is appropriate for such applications, and we develop it further in Section~\ref{apply}.

\section{Definitions and proof techniques}\label{theory}
\noindent
Let us now develop formal definitions and proof techniques for security and safety; these serve as background for Section \ref{sec:mpdyn}, where we present formal models and proofs for security and safety of ${\it NS}^{d+}$ with respect to ${\it IS}^{d+}$. 

%\subsection{Definitions}
Let $\preceq$ be a precongruence on processes %\footnote{A precongruence is a preorder that is closed under arbitrary contexts.} %By definition every context is monotonic in such a preorder.
and $\simeq$ % \:\triangleq\: \preceq \cap \preceq^{-1}$
be the associated congruence. A process $P$ under a context $\varphi$ is written as $\varphi[P]$. Contexts act as tests for behaviors, and $P \preceq Q$ means that any test that is passed by $P$ is passed by $Q$---in other words, ``$P$ has no more behaviors than $Q$". 

We describe an implementation as a binary relation $\mathcal R$ over processes, which relates specification instances to implementation instances. This relation conveniently generalizes a refinement map \cite{refmap}. %Next we define what we mean by a fully abstract implementation.
%Then $\mathcal R$ is fully abstract if it preserves and reflects $\preceq$. 

\begin{definition}[Full abstraction] An implementation $\mathcal R$ is fully abstract if it satisfies:% \textsc{(Preservation)} and \textsc{(Reflection)}:
$$\infer[(Preservation)]{}{\forall (P,Q) \in \mathcal R.~\forall (P',Q') \in \mathcal R.~~~P \preceq P' ~\Rightarrow~ Q \preceq Q'}$$
$$\infer[(Reflection)]{}{\forall (P,Q) \in \mathcal R.~\forall (P',Q') \in \mathcal R.~~~Q \preceq Q' ~\Rightarrow~ P \preceq P'}$$
%
%\[
%\left.
%\begin{array}{rl}
%\textsc{(Preservation)} & \\
%&~~~~~~~~~~~~~~~~~~~~~~~~~~\\
%\textsc{(Reflection)} & \forall (P,Q) \in \mathcal R.~\forall (P',Q') \in \mathcal R. \\
%&~~~~~~~~~~~~~~~~~~~~~~~~~~Q \preceq Q' ~\Rightarrow~ P \preceq P'
%\end{array}
%\right.
%\]
%$$\infer[(Preservation)]
%	{(P,Q) \in \mathcal R \\ (P',Q') \in \mathcal R}
%	{P \preceq P' ~\Rightarrow~ Q \preceq Q'}
%\qquad
%\infer[(Reflection)]
%	{(P,Q) \in \mathcal R \\ (P',Q') \in \mathcal R}
%	{Q \preceq Q' ~\Rightarrow~ P \preceq P'}
%$$
\end{definition}
\noindent
(\textsc{Preservation}) and (\textsc{Reflection}) are respectively soundness and completeness of the implementation under $\preceq$. Security only requires soundness.
\begin{definition}[\emph{cf.} Definition 2 [Security$\mbox{]}$] An implementation is secure if it satisfies \textsc{(Preservation)}.
\end{definition}
\noindent
Intuitively, a secure implementation does not \emph{introduce} any interesting behaviors---if $(P,Q)$ and $(P',Q')$ are in a secure $\mathcal R$ and $P$ has no more behaviors than $P'$, then $Q$ has no more behaviors than $Q'$. A fully abstract implementation moreover does not \emph{eliminate} any interesting behaviors.%; if $\mathcal R$ is fully abstract and $Q$ has no more behaviors than $Q'$,  then $P$ has no more behaviors than $P'$. 

Any subset of a secure implementation is secure. Security implies preservation of $\simeq$. 
%\begin{proposition} Let $\mathcal R$ be secure. Then it satisfies:
%$$\infer
%	{(P,Q) \in \mathcal R \\ (P',Q') \in \mathcal R}
%	{P \simeq P' ~\Rightarrow~ Q \simeq Q'}
%$$
%\end{proposition}
%\begin{proof} Suppose that $(P,Q) \in \mathcal R$ and $(P',Q') \in \mathcal R$. 
%\[
%\left.
%\begin{array}{rcll}
% P \simeq P' &  \Rightarrow & P \preceq P' ~\wedge~ P' \preceq P & \mbox{(by definition)} \\
%  & \Rightarrow & Q \preceq Q' ~\wedge~ Q' \preceq Q  & \mbox{(by assumption)} \\
%  &  \Rightarrow & Q \simeq Q'    & \mbox{(by definition)}
%\end{array}
%\right.
%\]
%\end{proof}
Finally, testing itself is trivially secure since $\preceq$ is closed under any context.
\begin{proposition}\label{trivprop} Let $\varphi$ be any context. Then $\{(P,\varphi[P])~|~P \in \mathcal W\}$ is secure for any set of processes $\mathcal W$.
\end{proposition}
%\begin{proof} $P \preceq P'~\Rightarrow~\varphi[P] \preceq \varphi[P']$ (by monotonicity).
%\end{proof}
\noindent
On the other hand, a context may eliminate some interesting behaviors by acting as a test for those behaviors. A fully abstract context does not; it merely \emph{translates} behaviors. 
\begin{definition}[Fully abstract context] A context $\varphi$ is fully abstract for a set of processes $\mathcal W$ if $\{(P,\varphi[P])~|~P \in \mathcal W\}$ is fully abstract.
\end{definition}
\noindent
A fully abstract context can be used as a wrapper to account for any benign differences between the implementation and the specification. An implementation is safe if it does not introduce any behaviors modulo such a wrapper.
\begin{definition}[\emph{cf.} Definition 1 [Safety$\mbox{]}$]\label{safetydef} An implementation $\mathcal R$ is safe if there exists a fully abstract context $\phi$ for the set of specification instances such that $\mathcal R$ satisfies:% \textsc{(Inclusion)}:
$$\infer[(Inclusion)]{}{\forall (P,Q) \in \mathcal R.~~Q \preceq \phi[P]}$$
%\[
%\left.
%\begin{array}{rl}
%\textsc{(Inclusion)} & 
%\end{array}
%\right.
%\]
%$$\infer[(Inclusion)]
%	{(P,Q) \in \mathcal R}
%	{Q \preceq \phi[P]}
%$$
\end{definition}
\noindent
Let us see why $\phi$ must be fully abstract in the definition. Suppose that it is not. Then for some $P$ and $P'$ we have $\phi[P] \preceq \phi[P']$ and $P \not\preceq P'$. Intuitively, $\phi$ ``covers up" the behaviors of $P$ that are not included in the behaviors of $P'$. 
%So there must be a context that $P$ passes but $P'$ does not, yet that context cannot be of the form $\_[\phi]$. Technically, $\phi$ is then a biased context that ``hides" any bad behavior that $\psi$ can test. %; in particular, $\phi$ acts as a biased context that hides potentially undesirable behaviors that $P$ shows that are not included in the behaviors of $P'$.
%Since $P$ leaks more information than $P'$, an implementation of $P$ is not necessarily an implementation of $P'$. 
Unfortunately, those behaviors may be unsafe. For instance, let $P'$ be a pi calculus process \cite{polypi} that does not contain public channels, and $\{P'\}$ be the set of specification instances---we consider any output on a public channel to be unsafe. Let $c$ be a public channel; let $P = \snd c {} P'$ and $\phi = \bullet~|~!\:\overline c \tup{}$. Then $P \not\preceq P'$ and $\phi[P] \preceq \phi[P']$, as required. But clearly $P$ is unsafe by our assumptions; yet $P \preceq \phi[P']$, so that by definition $\{(P',P)\}$ is safe! The definition therefore becomes meaningless. %Indeed, any relation that does not reflect $\simeq$ cannot be safe.

We now present some proof techniques. A direct proof of security requires mappings between subsets of $\preceq$. Those mappings may be difficult to define and manipulate. Instead a security proof may be built modularly by showing simulations, as in a safety proof. Such a proof requires simpler mappings between processes. %Next we show such a proof technique.
\begin{proposition}[Proof of security]\label{pf-fullabs}
Let $\phi$ and $\psi$ be contexts such that for all $(P,Q) \in \mathcal R$, 
	${Q \preceq \phi[P]}$,
	${P \preceq \psi[Q]}$, and
	${\phi[\psi[Q]] \preceq Q}$.
%$$\infer
%	{(P,Q) \in \mathcal R}
%	{Q \preceq \phi[P]}
%\qquad
%\infer
%	{(P,Q) \in \mathcal R}
%	{P \preceq \psi[Q]}
%\qquad
%\infer
%	{(\_,Q) \in \mathcal R}
%	{\phi[\psi[Q]] \preceq Q}
%$$
Then $\mathcal R$ is secure. 
\end{proposition}
\begin{proof} Suppose that $(P,Q) \in \mathcal R$, $P \preceq P'$, and $(P',Q') \in \mathcal R$. Then
$Q \preceq \phi[P]  \preceq \phi[P'] \preceq \phi[\psi[Q']] \preceq Q'$. 
\end{proof}
\noindent
Intuitively, $\mathcal R$ is secure if $\mathcal R$ and $\mathcal R^{-1}$ both satisfy \textsc{(Inclusion)}, and the witnessing contexts ``cancel" each other. A simple technique for proving full abstraction for contexts follows as a corollary. 
\begin{corollary}[Proof of full abstraction for contexts]\label{pf-pres} Let there be a context $\varphi^{-1}$ such that for all 
$P \in \mathcal W$,
	$\varphi^{-1}[\varphi[P]] \simeq P$. 
%$$\infer
%	{P \in \mathcal W}
%	{\varphi^{-1}[\varphi[P]] \simeq P}
%$$
Then $\varphi$ is a fully abstract context for $\mathcal W$.
\end{corollary}
\begin{proof} Take $\phi = \varphi^{-1}$ and $\psi = \varphi$ in the proposition above to show that $\{(\varphi[P],P)~|~P \in \mathcal W\}$ is secure. The converse follows by Proposition \ref{trivprop}.
\end{proof}

\paragraph{Theory for the applied pi calculus}

Let $a,b,\dots$ range over names, $u,v,\dots$ over names and variables, $M,N,\dots$ over terms, and $A,B,\dots$ over extended processes. Semantic relations include the binary relations $\equiv$, $\rightarrow$, and $\action\ell$ over extended processes (structural equivalence, reduction, and labeled transition); here labels $\ell$ are of the form $a(\seq M)$ or $\new{\seq u}\overline a\tup{\seq v}$ (where $a \notin \seq u$ and $\seq u \subseteq \seq v$). Both $\rightarrow$ and $\action\ell$ are closed under $\equiv$ and $\rightarrow$ is closed under arbitrary evaluation contexts. 

We recall some theory on may testing for applied pi calculus programs.
\begin{definition}[Barb] A barb $\downarrow_a$ is a predicate that tests possible output on $a$; we write $A \downarrow_a$ if $A \action{\new{\seq u}\overline a\tup{\seq v}} B$ for some $B$, $\seq v$, and $\seq u$. A weak barb $\Downarrow_a$ tests possible eventual output on $a$, \emph{i.e.}, $\Downarrow_a \:\triangleq\: \rightarrow^\star\downarrow_a$.
\end{definition}
\noindent
%
%$$\new{\seq n}\sigma \vdash M = N ~\stackrel\triangle=~ M\sigma = N\sigma~\wedge~ \seq n \cap (\fn(M) \cup \fn(N)) = \varnothing$$
%%
\begin{definition}[Frame]
Let $A$ be closed. Then we have $A \equiv (\nu\seq a)(\sigma ~|~P)$ for some $\seq a$, $\sigma$, and $P$ such that $\fv(\mathtt{rng}(\sigma)) \cup \fv(P) = \varnothing$; define $\mathtt{frame}(A) \equiv \new{\seq a}\sigma$.
\end{definition}
% and $\seq n \subseteq \fn(\sigma)$. 
\begin{definition}[Static equivalence] Let $A$ and $B$ be closed. Then $A$ is statically equivalent to $B$, written $A \approx_{\tt s} B$, if there exists $\seq a$, $\sigma$, and $\sigma'$ such that $\mathtt{frame}(A) \equiv \new{\seq a}\sigma$, $\mathtt{frame}(B) \equiv \new{\seq a}\sigma'$, $\dom(\sigma) = \dom(\sigma')$, and for all $M$ and $N$,
$$\{\seq a\} \cap (\fn(M) \cup \fn(N)) = \varnothing ~~\Rightarrow~~ M\sigma = N\sigma \Leftrightarrow M\sigma' = N\sigma'$$
\end{definition}
\begin{proposition} $A \approx_{\tt s} B$ if and only if $\mathtt{frame}(A) \simeq \mathtt{frame}(B)$.
\end{proposition}
\begin{proof} By induction on the structure of closing evaluation contexts.
\end{proof}

We can prove $\preceq$ by showing a simulation relation that approximates $\preceq$.
\begin{definition}[Simulation preorder]
Let $\preccurlyeq$ be the largest relation $\mathcal S$ such that for all $A$ and $B$, $(A,B) \in \mathcal S$ implies
\begin{itemize}
\item $A \approx_{\tt s} B$
\item $\forall A'.~~A \rightarrow A' ~\Rightarrow~ \exists B'.~~B \rightarrow^\star B' ~\wedge~ (A',B') \in \mathcal S$
\item $\forall A',\alpha.~~A \action\ell A' ~\Rightarrow~ \exists B'.~~B \rightarrow^\star\action\ell\rightarrow^\star B' ~\wedge~ (A',B') \in \mathcal S$
\end{itemize}
\end{definition}
\begin{proposition}[Proof of testing precongruence]\label{pf-prec} $\preccurlyeq \:\subseteq\: \preceq$. 
\end{proposition}

\section{Models and proofs for static access policies}\label{sec:mpstat}
We now present implementation and specification models and security proofs for static access policies. Models and proofs for dynamic access policies follow essentially the same routine, and are presented in the next section.

\subsection{Preliminaries}
We fix an equational theory $\Sigma$ with the following properties.
\begin{itemize}
\item $\Sigma$ includes a theory of natural numbers with symbols $0$ (zero), $\_+1$ (successor), and $\_\leq\_$ (less than or equal to).
%$$\forall k,k' \in \mathbb N.~~k' \neq k \Rightarrow \Sigma \vdash k' \neq k$$
%\item $\Sigma$ contains at least the constructors $\func{mac}(\cdot,\cdot)$, $\func{msg}(\cdot,\cdot)$, $\langle\cdot,\cdot\rangle$, $\func{proj}_1(\cdot)$, $\func{proj}_2(\cdot)$, $\func{auth}(\cdot,\cdot)$, $\func{ok}()$, and $\func{exec}(\cdot,\cdot)$.
\item $\Sigma$ includes a theory of finite tuples with symbols $\langle\_,\_\rangle$ (indexed concatenate) and $\_\:.\:\_$ (indexed project).
\item $\Sigma$ contains exactly one equation that involves the symbol $\func{mac}$, which is
$$\func{msg}(\func{mac}(x,y)) = x$$
\end{itemize}
Clients are identified by natural numbers; we fix a finite subset $\mathcal I$ of $\mathbb N$ and consider any user not identified in $\mathcal I$ to be dishonest. 

File-system code and other processes are conveniently modeled by parameterized process expressions, whose semantics are defined (recursively) by extending the usual semantic relations $\equiv$, $\rightarrow$, and $\action\ell$. 

\begin{figure}
\hspace{-0.55cm}
\fbox{\parbox{13.0cm}{\small
$$\infer[(Op Req)]{k \in \mathbb N}{\textsc{tfs}(F,\rho)^\eta
 \equiv \eta(\betac_k)(\op,x)\pause
\Req_k(F,\op,x)^\eta~|~\textsc{tfs}(F,\rho)^\eta}
\quad~~
%
%\infer[(Op Deny)]{\func{auth}(k,\op) = \func{no}}
%{\Req_k(\op,n)
%\rightarrow
%0}$$
%
\infer[(Op Ok)]{\mathtt{perm}(F,k,\op) = L}
{\Req_k(F,\op,M)
\rightarrow  
\Comm(L,\op,M)}
$$
$$\infer[(Op Exec)]{\func{exec}(L,\op,\rho) = \tup{N,\rho'}}
{\Comm(L,\op,M)~|~\textsc{tfs}(F,\rho)^\eta
\rightarrow 
\overline M\tup{N}~|~\textsc{tfs}(F,\rho')^\eta}
%\qquad
%%
%\infer[(Op Res Ret)]{}{\Ret(n,r) \equiv }
$$
%\line(1,0){370}
%
\fbox{\parbox{12.75cm}{
$$\infer[(Dummy Auth Req)]{j \in \mathbb N\backslash \mathcal I}{{\shortuparrow^\mathrm{TS}_\mathrm{NAS}}^\eta \equiv \eta(\alpha_{j})(\op,x)\pause \overline x \tup{\func{mac}(\tup{j,\op},K_?)}~|~{\shortuparrow^\mathrm{TS}_\mathrm{NAS}}^\eta}
%\qquad
%%
%\infer[(Dummy Auth Cap)]{}
%{\DAReq_{k}(\op,c) \equiv }
$$
$$\infer[(Dummy Exec Req)]{j \in \mathbb N\backslash \mathcal I}{{\shortuparrow^\mathrm{TS}_\mathrm{NAS}}^\eta \equiv \eta(\beta_{j})(\kappa,x)\pause \DReq(\kappa,x)^\eta~|~{\shortuparrow^\mathrm{TS}_\mathrm{NAS}}^\eta}
\qquad
\infer[(Dummy Op Req)]{\kappa = \func{mac}(\func{msg}(\kappa),K_?)\\\\ \func{msg}(\kappa) = \tup{j,\op} \\ j \in \mathbb N \backslash \mathcal I}
{\DReq(\kappa,M)^\eta \rightarrow \overline{\eta(\betac_j)} \tup{\op,M}}$$
}}
}}
\caption{A traditional file system with local access control}
\label{fig:ts-s}
%\vspace{-0.2cm}
\end{figure}

\begin{figure}
\hspace{-0.5cm}\fbox{\parbox{13.0cm}{\small
$$\infer[(Auth Req)]{k \in \mathbb N}{\textsc{nafs}(F,\rho)^\eta \equiv \eta(\alpha_{k})(\op,x)\pause \AReq_{k}(F,\op,x)~|~\textsc{nafs}(F,\rho)^\eta}
\qquad
\infer[(Auth Cap)]{\mathtt{cert}(F,k,\op) = \kappa}
{\AReq_{k}(F,\op,M) \rightarrow \overline M\tup{\kappa}}
%\qquad
%%
%\infer[(Auth Bad Cap)]{\func{auth}(k,\op) \neq \func{ok}}
%{\AReq_{k}(\op,c)^\eta \rightarrow \overline c\tup {\func{mac}(\tup{k,\op},\mathrm K_\bot)}}
$$
$$\infer[(Exec Req)]{k \in \mathbb N}{\textsc{nafs}(F,\rho)^\eta \equiv \eta(\beta_{k})(\kappa,x)\pause \Req(\kappa,x)~|~\textsc{nafs}(F,\rho)^\eta}
\qquad
\infer[(Op Ok)]{\mathtt{verif}(\kappa) = L \\ L \in \{\func{true},\func{false}\} \\\\ \func{msg}(\kappa) = \tup{\_,\op}}
{\Req(\kappa,M) \rightarrow  \Comm(L,\op,M)}$$
%
%$$\infer[(Op Not Ok)]{\kappa = \func{mac}(\func{msg}(\kappa),\mathrm K_\bot)}
%{\Req(\kappa,n)^\eta \rightarrow  \Comm(\func{false},\func{proj}_2(\func{msg}(\kappa)),n)^\eta}$$
%
%$$\infer[(Exec Deny)]{\kappa \neq \func{mac}(\func{msg}(\kappa),\mathrm K)}
%{\Req(\kappa,n) \rightarrow 0}$$
%%
$$\infer[(Op Exec)]{\func{exec}(L,\op,\rho) = \tup{N,\rho'}}
{\Comm(L,\op,M)~|~\textsc{nafs}(F,\rho)^\eta
\rightarrow 
\overline M\tup{N}~|~\textsc{nafs}(F,\rho')^\eta}
%\qquad
%%
%\infer[(Op Res Ret)]{}{\Ret(n,r)
%\action{\new x\overline n\tup x} \{r/x\}}
$$
%
%\line(1,0){370}
\fbox{\parbox{12.75cm}{
$$\infer[(Dummy Op Req)]{j \in \mathbb N\backslash \mathcal I}{{\shortuparrow^\mathrm{NAS}_\mathrm{TS}}^\eta \equiv \eta(\betac_{j})(\op,x)\pause \DReq_j(\op,x)^\eta~|~{\shortuparrow^\mathrm{NAS}_\mathrm{TS}}^\eta}
$$
$$\infer[(Dummy Auth \& Exec Req)]{}{\DReq_j(\op,M)^\eta \equiv \new c\overline{\eta(\alpha_{j})}\tup{\op,c} \pause c(\kappa)\pause \overline{\eta(\beta_j)}\tup{\kappa,M}}
$$
}}
}}
\caption{A network-attached file system with distributed access control}
   \label{fig:nas-s}
\vspace{-0.3cm}
\end{figure}

\subsection{Models}

Figures \ref{fig:ts-s} and \ref{fig:nas-s} show applied pi calculus models for the file systems under study. We ignore the rules in the inner boxes in these figures (labeled (\textsc{Dummy}...)) in a first reading.

Figure \ref{fig:ts-s} models a traditional file system (with local access control). The file system is parameterized by an access policy $F$, a store $\rho$, and a renaming $\eta$ of its default interface. That interface includes a channel $\betac_k$ for every $k \in \mathbb N$; intuitively, a user identified by $k$ may send operation requests on this channel. 

Processes $\Req_k(F,\op,n)$ and $\Comm(M,\op,n)$ denote internal states. 
In the equational theory $\func{auth}(F,k,\op) = \func{ok}$ means that user $k$ may access $\op$ under $F$, and $\func{exec}(L,\op,\rho) = \tup{N,\rho'}$ means that the execution of $\op$ on store $\rho$ under decision $L$ returns $N$ and store $\rho'$. Decisions are derived by $\mathtt{perm}(\_,\_,\_)$ as follows.
$$\infer
	{L = \func{true}\mbox{ if }\func{auth}(F,k,\op) = \func{ok}\mbox{, }=\func{false}\mbox{ otherwise}}
	{\mathtt{perm}(F,k,\op) \:\triangleq\: L}
$$
A traditional storage system may be described as 
$$(\nu_{i \in \mathcal I}\betac_i)(C~|~\textsc{ifs}(F,\rho))$$
Here $C$ is code run by honest users; the file-system exports the default interface (implicitly renamed by ``identity"), and channels associated with honest users are hidden from the context. The context may be arbitrary and is left implicit; in particular, channels associated with dishonest users are available to the context.

Figure \ref{fig:nas-s} models a network-attached file system (with distributed access control). As above, the file system is parameterized by an access policy $F$, a store $\rho$, and a renaming $\eta$ of its default interface. That interface includes channels $\alpha_k$ and $\beta_k$ for every $k \in \mathbb N$; intuitively, a user identified by $k$ may send authorization requests on $\alpha_k$ and execution requests on $\beta_k$. 

Processes $\AReq_k(F,\op,c)$, $\Req(\kappa,n)$, and $\Comm(M,\op,n)$ denote internal states. 
In the equational theory $\func{auth}(F,k,\op) = \func{ok}$ and $\func{exec}(L,\op,\rho) = \tup{N,\rho'}$ have the same meanings as above. Capabilities and decisions are derived by $\mathtt{cert}(\_,\_,\_)$ and  $\mathtt{verif}(\_)$ as follows.
$$\infer
	{a = K_{M\!D}\mbox{ if }\func{auth}(F,k,\op) = \func{ok}\mbox{, }=K'_M\mbox{ otherwise}}
	{\mathtt{cert}(F,k,\op) \:\triangleq\: \func{mac}(\tup{k,\op},a)}
$$
$$\infer
	{L = \func{true}\mbox{ if }\kappa = \func{mac}(\func{msg}(\kappa),K_{M\!D})\mbox{, }=\func{false}\mbox{ if }\kappa = \func{mac}(\func{msg}(\kappa),K'_M)}
	{\mathtt{verif}(\kappa) \:\triangleq\: L}
$$
A network-attached storage system may be described as 
$$(\nu_{i \in \mathcal I}\alpha_i\beta_i)(C~|~\new{K_{M\!D}K'_M}\textsc{nafs}(F,\rho))$$
As above, $C$ is code run by honest users; the file-system exports the default interface and hides the keys that authenticate capabilities. Channels associated with honest users are hidden from the context. The context may be arbitrary and is left implicit; in particular, channels associated with dishonest users are available to the context.

\subsection{Proofs of security}
We prove that the implementation is secure, safe, and fully abstract with respect to the specification. We begin by outlining the proofs, and then present details.
\subsubsection{Outline}
Let $F$, $\rho$, and $C$ range over access policies, stores, and code for honest users that are ``wellformed" in the implementation. Let $\lceil\_\rceil$ abstract such $F$, $\rho$, and $C$ in the specification. We define
$$\mathcal R = \bigcup_{F,\rho,C}\{(\nu_{i \in \mathcal I}\betac_i)(\lceil C\rceil~|~\textsc{tfs}(\lceil F\rceil,\lceil\rho\rceil))~~, ~~(\nu_{i \in \mathcal I}\alpha_i\beta_i)(C~|~\new{K_{M\!D}K'_M}\textsc{nafs}(F,\rho))\}$$
We prove that $\mathcal R$ is secure by showing contexts $\phi$ and $\psi$ such that:
\begin{figure}
\hspace{-0.6cm}\fbox{\parbox{13.2cm}{\small
%\vspace{-0.3cm}
%$$\forall P,A.~~(\exists Q.~~\lceil P\rceil_A = Q)~ \Rightarrow ~$$ %\cup \{\betac_k~|~k \in \mathbb N\}$$
$$
\infer{\fn(M) \cap (\mathcal A \cup \{\alpha_j,\beta_j~|~j \in \mathbb N \backslash \mathcal I\}) = \varnothing}{\lceil M\rceil = M} \qquad
\infer{\lceil P\rceil_\Gamma = Q \\ \Gamma \supseteq \{\alpha_j,\beta_j~|~j \in \mathbb N \backslash \mathcal I\}}{\lceil P\rceil = Q}$$
$$
\infer{\dom(\Gamma) \supseteq \mathcal A}{\lceil 0\rceil_\Gamma = 0} \qquad
\infer{n \notin \dom(\Gamma)}{\lceil \new{n}{P}\rceil_\Gamma  =   \new{n}{\lceil P\rceil_\Gamma}} \qquad
\lceil P~|~Q \rceil_\Gamma   =   \lceil P\rceil_\Gamma~|~\lceil Q\rceil_\Gamma
\qquad \lceil !P\rceil_\Gamma   =   \:!\lceil P\rceil_\Gamma
$$
$$
\infer{\fnv(u,\seq x) \cap \dom(\Gamma) = \varnothing}{\lceil \rcv{u}{\seq x}{P}\rceil_\Gamma   =   \rcv{u}{\seq x}{\lceil P\rceil_\Gamma}} \qquad
\infer{\fnv(u,\seq M)\cap \dom(\Gamma) =\varnothing}{\lceil \snd{u}{\seq M}{P}\rceil  = \snd{u}{\seq M}{\lceil P\rceil}}$$
$$
\infer{\fnv(M,N) \cap \dom(\Gamma) = \varnothing}{\lceil \cond{M}{N}{P}{Q}\rceil_\Gamma   =   \cond{M}{N}{\lceil P \rceil_\Gamma}{\lceil Q \rceil_\Gamma}} 
$$
$$
\infer{i \in \mathcal I \\\\ \fnv(c,x) \cap \dom(\Gamma) = \varnothing \\ c \notin \fn(P)}{\lceil \new c\overline{\alpha_i}\tup{\op,c}\pause c(x)\pause P\rceil_\Gamma   =
  \lceil P\rceil_{\Gamma, x : \mathtt{Cert}(i,\op)}} \qquad
  \infer{\{i,i'\} \subseteq \mathcal I \\ \Gamma(x) = \mathtt{Cert}(i',\op) \\\\ \fnv(\op,M) \cap \dom(\Gamma) = \varnothing}{\lceil \overline{\beta_i}\tup{x,M}\pause P\rceil_\Gamma = \overline{\betac_{i'}}\tup{\op,M}\pause \lceil P\rceil_\Gamma}$$
}}
\caption{Abstraction function}
   \label{fig:abs-s}
\end{figure}
\begin{lemma}\label{static-lemma} For any $F$, $\rho$, and $C$,
\begin{enumerate}
\item $(\nu_{i \in \mathcal I}\alpha_i\beta_i)(C~|~\new{K_{M\!D}K'_M}\textsc{nafs}(F,\rho)) ~\preceq~ \phi[(\nu_{i \in \mathcal I}\betac_i)(\lceil C\rceil~|~\textsc{tfs}(\lceil F\rceil,\lceil \rho\rceil))] $
\item $(\nu_{i \in \mathcal I}\betac_i)(\lceil C\rceil~|~\textsc{tfs}(\lceil F\rceil,\lceil \rho\rceil)) ~\preceq ~ \psi[(\nu_{i \in \mathcal I}\alpha_i\beta_i)(C~|~\new{K_{M\!D}K'_M}\textsc{nafs}(F,\rho))] $
\item $\phi[\psi[(\nu_{i \in \mathcal I}\alpha_i\beta_i)(C~|~\new{K_{M\!D}K'_M}\textsc{nafs}(F,\rho))]] $
\par $~~~~~~~~~~~~~~~~~~~~~~~~~~~~~~~~~~~~~~~~~~~~~~~~~~~~~~~\preceq ~ (\nu_{i \in \mathcal I}\alpha_i\beta_i)(C~|~\new{K_{M\!D}K'_M}\textsc{nafs}(F,\rho))$
\end{enumerate}
\end{lemma}
Proposition \ref{pf-fullabs} then applies. Moreover we show:
\begin{lemma}\label{static-safetylemma} For any $F$, $\rho$, and $C$, 
$$\psi[\phi[(\nu_{i \in \mathcal I}\betac_i)(\lceil C\rceil~|~\textsc{tfs}(\lceil F\rceil,\lceil \rho\rceil))]] ~\preceq ~ (\nu_{i \in \mathcal I}\betac_i)(\lceil C\rceil~|~\textsc{tfs}(\lceil F\rceil,\lceil \rho\rceil))$$
\end{lemma}
Now $\mathcal R^{-1}$ is secure by Proposition \ref{pf-fullabs}. Thus $\mathcal R$ is proved fully abstract. Moreover Lemmas \ref{static-lemma}.1--2 already imply the converse of Lemma \ref{static-safetylemma}; so $\phi$ is a fully abstract context by Corollary \ref{pf-pres} (taking $\phi^{-1} = \psi$). Thus $\mathcal R$ is proved safe. 

We now revisit Figures \ref{fig:ts-s} and \ref{fig:nas-s} and focus on the rules in the inner boxes. Those rules define processes $\shortuparrow^\mathrm{TS}_\mathrm{NAS}$ and $\shortuparrow^\mathrm{NAS}_\mathrm{TS}$. Intuitively, these processes translate public requests from ${\it NAS}^s$ to ${\it TS}^s$ and from ${\it TS}^s$ to ${\it NAS}^s$. Let $\seq a_\mathrm{TS}$ and $\seq a_\mathrm{NAS}$ include the public interfaces of ${\it TS}^s$ and ${\it NAS}^s$. We define 
\begin{eqnarray*}
\phi & = & \new{\seq a_\mathrm{TS}} (\bullet~|~\shortuparrow^\mathrm{TS}_\mathrm{NAS}) \\
\psi & = & \new{\seq a_\mathrm{NAS}} (\bullet~|~\shortuparrow^\mathrm{NAS}_\mathrm{TS}) 
\end{eqnarray*}
The abstraction function $\lceil\_\rceil$ is shown in Figure \ref{fig:abs-s}. Here $\mathcal A$ contains special names whose uses in well-formed code are either disciplined or forbidden. 
$$\mathcal A \triangleq  \{\alpha_i,\beta_i~|~i \in \mathcal I\} \cup \{{\alpha_j}_?, {\beta_j}_?,{\betac_j}_?~|~j \in \mathbb N\backslash \mathcal I\} \cup \{K_{M\!D},K'_M,K_?\}$$
The names in $\{{\alpha_j}_?, {\beta_j}_?,{\betac_j}_?~|~j \in \mathbb N\backslash \mathcal I\} \cup \{K_?\}$ are invented to simplify proofs below.
\begin{figure}
\hspace{-0.5cm}\fbox{\parbox{13.0cm}{\small
$$\infer
	{\fn(\kappa,M) \cap \mathcal A = \varnothing}
	{\Req(\kappa,M) ~~\mathcal S'^F_1~~ \DReq(\kappa,M)^{\eta_2}}
\qquad
\infer
	{k \in \mathbb N \\ \fn(\op,M) \cap \mathcal A = \varnothing}
	{\Req(\mathtt{cert}(F,k,\op),M) ~~\mathcal S'^F_1~~ \Req_k(F,\op,M)}
$$
$$\infer
	{\fn(L,\op,M) \cap \mathcal A = \varnothing}
	{\Comm(L,\op,M) ~~\mathcal S'^F_1~~ \Comm(L,\op,M)}
\qquad
\infer
	{j \in \mathbb N \backslash \mathcal I \\ \fn(L,\op,M) \cap \mathcal A = \varnothing}
	{\AReq_j(F,\op,M) ~~\mathcal S'^F_1~~ \overline{M}\tup{\func{mac}(\tup{j,\op},\mathrm K_?)}}
$$
$$\inferc{File systems}
	{\forall r \in \mathcal L.~~P_r~~\mathcal S'^F_1~~Q_r \\ \fn(\rho) \cap \mathcal A = \varnothing}
	{\textsc{nafs}(F,\rho)~|~\Pi_{r \in \mathcal L} P_r ~~\mathcal S_1^F~~ \textsc{tfs}(F,\rho)^{\eta_2}~|~\Pi_{r \in \mathcal L} Q_r}
$$
$$\inferc{Honest users}
	{\forall x.~~x \in \mathtt{dom}(\sigma)~\Rightarrow~\exists i \in \mathcal I,\op.~~\Gamma(x) = \mathtt{Cert}(i,\op) ~\wedge ~\sigma(x) = \mathtt{cert}(F,i,\op)} 
	%\\\forall x.~~x \in \mathtt{dom}(\sigma')~\Rightarrow~x \notin \dom(\Gamma)~\wedge~\exists j \in \mathbb N\backslash\mathcal I,\op.~~\sigma'(x) = \mathtt{cert}(F,j,\op)}
%	{C\sigma'\sigma ~~\mathcal S_2^F~~ \eta_3(\lceil C\rceil_\Gamma\sigma')}
	{C\sigma ~~\mathcal S_2^{\Gamma,F}~~ \lceil C\rceil_\Gamma}
$$
$$\infer
	{i \in \mathcal I \\ P~~\mathcal S_2^{\Gamma,F}~~Q \\ \Gamma(x) = \mathtt{Cert}(i,\op)}
	{(\nu c)(c(x)\pause P~|~\AReq_i(F,\op,c)) ~~\mathcal S_3^F~~ Q}
\qquad
\infer
	{i \in \mathcal I \\ P~~\mathcal S_2^{\Gamma,F}~~Q \\ \Gamma(x) = \mathtt{Cert}(i,\op)}
	{(\nu c)(c(x)\pause P~|~\overline c\tup{\mathtt{cert}(F,i,\op)}) ~~\mathcal S_3^F~~ Q}
$$
$$\inferc{Trusted code}
	{P~~\mathcal S_1^F~~Q \\ P'~~\mathcal S_2^{\Gamma,F}~~Q' \\ \forall r \in \mathcal L.~~P_r~~\mathcal S_3^F~~Q_r}
	{(\nu_{i \in \mathcal I}\alpha_i\beta_i)(P~|~P'~|~\Pi_{r \in \mathcal L} P_r) ~~\mathcal S'^F~~ (\nu_{i \in \mathcal I}\betac_i)(Q~|~Q'~|~\Pi_{r \in \mathcal L} Q_r)}
$$
$$\inferc{System code}
	{P~~\mathcal S'^F~~Q \\ \forall x,N.~~(\exists \sigma'.~~\sigma \equiv \subs N x~|~\sigma')~\Rightarrow~N:_F \mathtt{Export}}
	{(\nu \seq n)(\nu K_{M\!D}K'_M)(\sigma~|~P)~~\mathcal S~~(\nu \seq n)(\nu K_?)(\eta_3(\sigma)~|~(\nu_{j \in \mathbb N\backslash\mathcal I}{\betac_j}_?)(Q~|~{\shortuparrow^\mathrm{TS}_\mathrm{NAS}}^{\eta_2}))}
$$
%$$\inferc{tfs-nafs-Final-CUT}
%	{P~~\mathcal T'~~Q \\ \forall x,M.~~(\exists \sigma'.~~\sigma \equiv \subs M x~|~\sigma')~\Rightarrow~M: \mathtt{Export}}
%	{(\nu \seq n)(\nu \mathrm K\mathrm K_\bot)(\sigma~|~(\nu_{i \in \mathcal I}\alpha_i\beta_i)~P)~~\mathcal T~~(\nu \seq n)(\nu \mathrm K_?)(\eta_3(\sigma)~|~(\nu_{j \in \mathbb N\backslash\mathcal I}\betac_j)((\nu_{i \in \mathcal I}\betac_i)~Q~|~\bowtie\!\textsc I))}
%$$
}}
\caption{Simulation relation for Lemma \ref{static-lemma}.1 ($\_ \preccurlyeq \phi[\_]$)}
\label{fig:simreln-1-s}
\end{figure}
\begin{figure}
\hspace{-0.4cm}\fbox{\parbox{12.8cm}{\small
$$\infer
	{k \in \mathbb N \\ \fn(\op,M) \cap \mathcal A = \varnothing}
	{\Req_k(F,\op,M) ~~\mathcal T'_1~~ \Req(\mathtt{cert}(F,k,\op),M)}
\qquad
\infer
	{\fn(L,\op,M) \cap \mathcal A = \varnothing}
	{\Comm(L,\op,M) ~~\mathcal T'_1~~ \Comm(L,\op,M)}
$$
$$\inferc{File systems}
	{\forall r \in \mathcal L.~~P_r~~\mathcal T'_1~~Q_r \\ \fn(\rho) \cap \mathcal A = \varnothing}
	{\textsc{tfs}(F,\rho)~|~\Pi_{r \in \mathcal L} P_r ~~\mathcal T_1^F~~ \textsc{nafs}(F,\rho)^{\eta_1}~|~\Pi_{r \in \mathcal L} Q_r }
$$
$$\inferc{Honest users}
	{\forall x.~~x \in \mathtt{dom}(\sigma)~\Rightarrow~\exists i \in \mathcal I,\op.~~\Gamma(x) = \mathtt{Cert}(i,\op) ~\wedge ~\sigma(x) = \mathtt{cert}(F,i,\op)}
	{ \lceil C\rceil_\Gamma~~\mathcal T_2^F~~ C\sigma}
$$
$$\inferc{System code}
	{P_1~~\mathcal T_1^F~~Q_1 \\ P_2~~\mathcal T_2^F~~Q_2  \\\\
	P = (\nu_{i \in \mathcal I}\betac_i)(P_1~|~P_2) \\ Q = (\nu_{i \in \mathcal I}\alpha_i\beta_i)(\nu K_{M\!D}K'_M)(Q_1~|~Q_2)}
	{(\nu \seq n)(\sigma~|~P) ~~\mathcal T~~ (\nu \seq n)(\sigma~|~(\nu_{j \in \mathbb N\backslash\mathcal I}{\alpha_j}_?{\beta_j}_?)(Q~|~{\shortuparrow^\mathrm{NAS}_\mathrm{TS}}^{\eta_1}))}
$$
%$$\inferc{tfs-nafs-Final-CUT}
%	{P_1~~\mathcal S_1~~Q_1 \\ P_2~~\mathcal S_2~~Q_2}
%	{(\nu \seq n)(\sigma~|~(\nu_{i \in \mathcal I}\betac_i)(P_1~|~P_2)) ~~\mathcal S~~ (\nu \seq n)(\sigma~|~(\nu_{j \in \mathbb N\backslash\mathcal I}\alpha_j\beta_j)((\nu_{i \in \mathcal I}\alpha_i\beta_i)(\new{K_{M\!D}K'_M}Q_1~|~Q_2)~|~{\shortuparrow^\mathrm{NAS}_\mathrm{TS}}))}
%$$
}}
\caption{Simulation relation for Lemma \ref{static-lemma}.2 ($\_ \preccurlyeq \psi[\_]$)}
\label{fig:simreln-2-s}
\end{figure}
\begin{figure}
\hspace{-0.5cm}\fbox{\parbox{13.0cm}{\small
$$\infer
	{\fn(\kappa,M) \cap \mathcal A = \varnothing}
	{\DReq(\kappa,M)^{\eta_2} ~~\mathcal U'^F_1~~ \Req(\kappa,M)}
\qquad
\infer
	{j \in \mathbb N \backslash I \\ \fn(\op,M) \cap \mathcal A = \varnothing}
	{\overline{{\betac_j}_?}\tup{\op,M} ~~\mathcal U'^F_1~~ \Req(\mathtt{cert}(F,j,\op) ,M)}
$$
$$\infer
	{j \in \mathbb N \backslash I \\ \fn(\op,M) \cap \mathcal A = \varnothing}
	{\DReq_j(\op,M)^{\eta_1\oplus\eta_2} ~~\mathcal U'^F_1~~ \Req(\mathtt{cert}(F,j,\op) ,M)}
$$
$$\infer
	{j \in \mathbb N \backslash I \\ \fn(\op,M) \cap \mathcal A = \varnothing}
	{(\nu c)(c(x)\pause \overline{{\beta_j}_?}\tup{x,M}~|~\AReq_j(F,\op,c)) ~~\mathcal U'^F_1~~ \Req(\mathtt{cert}(F,j,\op),M)}
$$
$$\infer
	{j \in \mathbb N \backslash I \\ \fn(\op,M) \cap \mathcal A = \varnothing}
	{(\nu c)(c(x)\pause \overline{{\beta_j}_?}\tup{x,M}~|~\overline c\tup{\mathtt{cert}(F,j,\op)}) ~~\mathcal U'^F_1~~ \Req(\mathtt{cert}(F,j,\op),M)}
$$
$$\infer
	{j \in \mathbb N \backslash I \\ \fn(\op,M) \cap \mathcal A = \varnothing}
	{\overline{{\beta_j}_?}\tup{\mathtt{cert}(F,j,\op),M} ~~\mathcal U'^F_1~~ \Req(\mathtt{cert}(F,j,\op) ,M)}
$$
$$\infer
	{j \in \mathbb N \backslash I \\ \fn(\op,M) \cap \mathcal A = \varnothing}
	{\Req(\mathtt{cert}(F,j,\op),M) ~~\mathcal U'^F_1~~ \Req(\mathtt{cert}(F,j,\op),M)}
$$
$$\infer
	{\fn(L,\op,M) \cap \mathcal A = \varnothing}
	{\Comm(L,\op,M) ~~\mathcal U'^F_1~~ \Comm(L,\op,M)}
\qquad
\infer
	{j \in \mathbb N \backslash I \\ \fn(\op,M) \cap \mathcal A = \varnothing}
	{\overline{M}\tup{\func{mac}(\tup{j,\op},\mathrm K_?)}~~\mathcal U'^F_1~~ \AReq_j(F,\op,M)}
$$
$$\inferc{File systems}
	{\forall r \in \mathcal L.~~P_r~~\mathcal U'^F_1~~Q_r \\ \fn(\rho) \cap \mathcal A = \varnothing} 
	{{\shortuparrow^\mathrm{TS}_\mathrm{NAS}}^{\eta_2}~|~{\shortuparrow^\mathrm{NAS}_\mathrm{TS}}^{\eta_1\oplus\eta_2} ~|~\textsc{nafs}(F,\rho)^{\eta_1}~|~\Pi_{r \in \mathcal L} P_r ~~\mathcal U_1^F~~ \textsc{nafs}(F,\rho)~|~\Pi_{r \in \mathcal L} Q_r}
$$
$$\inferc{Honest users}
	{\lceil C\rceil_\Gamma = C^\circ \\ %\forall x.~~x \in \dom(\sigma)~\Rightarrow~x \notin \dom(\Gamma) ~\wedge~\exists j \in \mathbb N\backslash \mathcal I,\op.~~\sigma(x) = \mathtt{cert}(F,j,\op) \\\\
	\forall x.~~x \in \dom(\sigma)~\Rightarrow~\exists i \in \mathcal I,\op.~~\Gamma(x) = \mathtt{Cert}(i,\op) ~\wedge ~\sigma(x) = \mathtt{cert}(F,i,\op)}
	{C\sigma ~~\mathcal U_2^{\Gamma,F}~~ C\sigma}
$$
$$\infer
	{i \in \mathcal I \\ P~~\mathcal U_2^{\Gamma,F}~~Q \\ \Gamma(x) = \mathtt{Cert}(i,\op)}
	{(\nu c)(c(x)\pause P~|~\AReq_i(F,\op,c)) ~~\mathcal U_3^F~~ (\nu c)(c(x)\pause Q~|~\AReq_i(F,\op,c))}
$$
$$\inferc{Trusted code}
	{P~~\mathcal U_1^F~~Q \\ P'~~\mathcal U_2^F~~Q' \\ \forall r \in \mathcal L.~~P_r~~\mathcal U_3^F~~Q_r}
	{(\nu_{i \in \mathcal I}\alpha_i\beta_i)(\nu K_{M\!D}K'_M)(P~|~P'~|~\Pi_{r \in \mathcal L} P_r) ~~\mathcal U'^F~~ (\nu_{i \in \mathcal I}\alpha_i\beta_i)(Q~|~Q'~|~\Pi_{r \in \mathcal L} Q_r)}
$$
%$$\inferc{tfs-nafs}
%	{P~~\mathcal U'~~Q \\ \forall x,M.~~(\exists \sigma'.~~\sigma \equiv \subs M x~|~\sigma')~\Rightarrow~M \notin \{\mathrm K,\mathrm K_\bot\}}
%	{(\nu \seq n)(\nu \mathrm K\mathrm K_\bot)(\sigma~|~(\nu_{i \in \mathcal I}\alpha_i\beta_i)~P)~~\mathcal U~~(\nu \seq n)(\nu \mathrm K_?)(\eta_3(\sigma)~|~(\nu_{j \in \mathbb N\backslash\mathcal I}\betac_j)(\nu\mathrm K\mathrm K_\bot)((\nu_{i \in \mathcal I}\betac_i)~Q~|~\bowtie\!\textsc I))}
%$$
$$\inferc{System code}
	{P~~\mathcal U'^F~~Q \\ \forall x,N.~~(\exists \sigma'.~~\sigma \equiv \subs N x~|~\sigma')~\Rightarrow~N:_F\mathtt{Export}}
	{(\nu \seq n)(\nu K_?)(\eta_3(\sigma)~|~(\nu_{j \in\mathbb N\backslash\mathcal I}{\betac_j}_?{\alpha_j}_?{\beta_j}_?)~P)~~\mathcal U~~(\nu \seq n)(\nu K_{M\!D}K'_M)(\sigma~|~Q)}
$$
}}
\caption{Simulation relation for Lemma \ref{static-lemma}.3 ($\phi[\psi[\_]] \preccurlyeq \_$)}
\label{fig:simreln-3-s}
\end{figure}

\begin{figure}[h]
%\vspace{-0.5cm}
\hspace{-0.5cm}\fbox{\parbox{13.0cm}{\small
$$\infer
	{j \in \mathbb N \backslash \mathcal I \\ \fn(\op,M) \cap \mathcal A = \varnothing}
	{\DReq_j(\op,M) ~~\mathcal V'^{F}_1~~ \Req_j(\op,M)}
\qquad
\infer
	{j \in \mathbb N \backslash \mathcal I \\ \fn(\op,\tau,M) \cap \mathcal A = \varnothing \\\\ N = \func{mac}(\tup{j,\op},K_?)}
	{(\nu c)(c(\kappa)\pause \overline{{\beta_j}_?}\tup{\kappa,M}~|~\overline c\tup{N}) ~~\mathcal V'^{F}_1~~ \Req_j(\op,M)}
$$
$$\infer
	{j \in \mathbb N \backslash \mathcal I \\ \fn(\op,M) \cap \mathcal A = \varnothing\\\\
	N = \func{mac}(\tup{j,\op},K_?)  \\ L = \mathtt{perm}(F,j,\op)}
	{\overline{{\beta_j}_?}\tup{N,M} ~~\mathcal V'^{F}_1~~ \Comm(L,\op,M)}
\quad
\infer
	{j \in \mathbb N \backslash \mathcal I \\ \fn(\op,M) \cap \mathcal A = \varnothing \\\\
	N = \func{mac}(\tup{j,\op},K_?)  \\ L = \mathtt{perm}(F,j,\op)}
	{\DReq(N,M)^{\eta_1\oplus\eta_2} ~~\mathcal V'^{F}_1~~ \Comm(L,\op,M)}
$$
$$\infer
	{j \in \mathbb N \backslash \mathcal I \\ \fn(\op,M) \cap \mathcal A = \varnothing  \\ L = \mathtt{perm}(F,j,\op)}
	{\overline{{\betac_j}_?}\tup{\op,M} ~~\mathcal V'^{F}_1~~ \Comm(L,\op,M)}
$$
$$\infer
	{\fn(\op,M) \cap \mathcal A = \varnothing}
	{\Comm(L,\op,M) ~~\mathcal V'^{F}_1~~ \Comm(L,\op,M)}
\qquad
\infer
	{\fn(\adm,M) \cap \mathcal A = \varnothing}
	{\PReq_k(\adm,M) ~~\mathcal V'^{F}_1~~ \PReq_k(\adm,M)}
$$
%\qquad
%\infer
%	{j \in \mathbb N \backslash \mathcal I \\ \fn(M) \cap \mathcal A = \varnothing}
%	{\new c\overline{{\alpha_j}_?}\tup{M,c}\pause c(y)\pause \overline M\tup{\func{msg}(y).3}~~\mathcal V'^{F,\clk}_1~~ \CReq(M)}
%$$
%$$\infer
%	{j \in \mathbb N \backslash \mathcal I \\ \fn(M) \cap \mathcal A = \varnothing}
%	{(\nu c)(c(y)\pause \overline M\tup{\func{msg}(y).3}~|~(\nu m)~\overline{{\alphac_j}_?}\tup{m}\pause m(x)\pause  \overline c\tup{\func{mac}(\tup{j,M,x},K_?)})
%~~\mathcal V'^{F,\clk}_1~~ \CReq(M)}
%$$
%$$\infer
%	{j \in \mathbb N \backslash \mathcal I \\ \fn(M) \cap \mathcal A = \varnothing}
%	{(\nu c)(c(y)\pause \overline M\tup{\func{msg}(y).3}~|~(\nu m)(m(x)\pause  \overline c\tup{\func{mac}(\tup{j,M,x},K_?)}
%~|~\CReq(M)))~~\mathcal V'^{F,\clk}_1~~ \CReq(M)}
%$$
%$$\infer
%	{j \in \mathbb N \backslash \mathcal I \\ \fn(M) \cap \mathcal A = \varnothing \\ \clk' \leq \clk}
%	{(\nu c)(c(y)\pause \overline M\tup{\func{msg}(y).3}~|~(\nu m)(m(x)\pause  \overline c\tup{\func{mac}(\tup{j,M,x},K_?)}
%~|~\overline m\tup{\clk'}))~~\mathcal V'^{F,\clk}_1~~ \overline m\tup{\clk'}}
%$$
%$$\infer
%	{j \in \mathbb N \backslash \mathcal I \\ \fn(M) \cap \mathcal A = \varnothing}
%	{(\nu c)(c(y)\pause \overline M\tup{\func{msg}(y).3}~|~\overline c\tup{\func{mac}(\tup{j,M,\clk'},K_?)})~~\mathcal V'^{F,\clk}_1~~ \overline M\tup{\clk'}}
%$$
%$$\inferc{tfs-nafs}
%	{}
%	{\overline c\tup{\func{mac}(\tup{\tup{j,\eta_3(\op)},\clk+1},\mathrm K_?)}~~\mathcal V'_1~~ \AReq_j(\op,c)}
%$$
$$\inferc{file systems}
	{\forall r \in \mathcal L.~~P_r~~\mathcal V'^{F}_1~~Q_r \\ \fn(\rho) \cap \mathcal A = \varnothing}
	{{\shortuparrow^\mathrm{NAS}_\mathrm{TS}}^{\eta_1}~|~{\shortuparrow^\mathrm{TS}_\mathrm{NAS}}^{\eta_1\oplus\eta_2}~|~\textsc{tfs}(F,\rho)^{\eta_2}~|~\Pi_{r \in \mathcal L} P_r ~~\mathcal V^{F,\clk}_1~~ \textsc{tfs}(F,\rho)~|~\Pi_{r \in \mathcal L} Q_r}
$$
$$\inferc{honest users}
	{}
	{\lceil C\rceil_\Gamma ~~\mathcal V^{F}_2~~ \lceil C\rceil_\Gamma}
\qquad
%$$\infer
%	{i \in \mathcal I \\ \Gamma(x) = \mathtt{Cert}(i,\op) \\ P~~\mathcal V^{\Gamma,F,\clk}_2~~Q}
%	{(\nu c)(c(x)\pause P~|~\CReq(c)) ~~\mathcal V^{F,\clk}_3~~ (\nu c)(c(x)\pause Q~|~\CReq(c))}
%$$
\inferc{system code}
	{P~~\mathcal V^{F}_1~~Q \\ P'~~\mathcal V^{F}_2~~Q' \\\\ %\forall r \in \mathcal L.~~P_\ell~~\mathcal V^{F,\clk}_3~~Q_\ell \\\\
	P'' = (\nu_{i \in \mathcal I}\betac_i)(\nu K_?)(P~|~P') \\ Q'' = (\nu_{i \in \mathcal I}\betac_i)(Q~|~Q')}
	{(\nu \seq n)(\sigma~|~(\nu_{j \in\mathbb N\backslash\mathcal I}{\betac_j}_?{\alpha_j}_?{\beta_j}_?)~P'')~~\mathcal V~~(\nu \seq n)(\sigma~|~Q'')}
$$
}}
\caption{Simulation relation for Lemma \ref{static-safetylemma} ($\psi[\phi[\_]] \preccurlyeq \_$)}
\label{fig:simreln-4-s}
\end{figure}

\subsubsection{Simulation relations}

Figures \ref{fig:simreln-1-s}, \ref{fig:simreln-2-s}, and \ref{fig:simreln-3-s} show simulation relations for Lemma \ref{static-lemma}.1--3. All these relations are closed under $\equiv$. Here $\eta_1$ and $\eta_2$ rename the public interfaces of ${\it NAS}^s$ and ${\it TS}^s$ and $\eta_3$ renames the private authentication keys $K_{M\!D}$ and $K'_M$. 
\begin{eqnarray*}
\eta_1 &\triangleq& [\alpha_j \mapsto {\alpha_j}_?,\beta_j \mapsto {\beta_j}_?~|~j \in \mathbb N\backslash \mathcal I]\\
\eta_2 &\triangleq& [\betac_j \mapsto {\betac_j}_?~|~j \in \mathbb N\backslash \mathcal I]\\
\eta_3 &\triangleq& [a \mapsto K_?~|~a \in \{K_{M\!D},K'_M\}]
\end{eqnarray*}
These renamings map to names in $\mathcal A$ that do not occur in wellformed code (see Figure~\ref{fig:abs-s}). 
In particular, the purpose of $\eta_1$ and $\eta_2$ is to rename some public channels to fresh ones that can be hidden by restriction in $\psi$ and $\phi$. (A similar purpose is served by quantification in logic.) Hiding those names strengthens Lemmas \ref{static-lemma}.1--2 while not affecting their proofs; but more importantly, the restrictions are required to prove Lemma \ref{static-lemma}.3. Further the purpose of $\eta_3$ is to abstract terms that may be available to contexts. Such terms must be of type $\mathtt{Export}$; intuitively, $K_{M\!D}$ and $K'_M$ may appear only as authentication keys in capabilities issued to dishonest users.
$$\infer
	{N = N'\sigma \\ \{K_{M\!D},K'_M,K_?\}\cap \fn(N') = \varnothing \\ \forall L \in \mathtt{rng}(\sigma).~~\exists j \in \mathbb N \backslash \mathcal I,\op.~~L = \mathtt{cert}(F,j,\op)~\wedge~\op :_F \mathtt{Export}}
	{N :_F \mathtt{Export}}
$$
We show that term abstraction preserves equivalence in the equational theory. 
\begin{lemma} Suppose that $M:_F \mathtt{Export}$ and $N:_F \mathtt{Export}$. Then $M = N$ iff $\eta_3(M) = \eta_3(N)$.
\end{lemma}
This lemma is required to show static equivalence in proofs of soundness for the relations $\mathcal S$, $\mathcal T$, and $\mathcal U$ in Figures \ref{fig:simreln-1-s}, \ref{fig:simreln-2-s}, and \ref{fig:simreln-3-s}, which in turn lead to Lemma \ref{static-lemma}. We prove that those relations are included in the simulation preorder.
\begin{lemma} $\mathcal S \subseteq\: \preccurlyeq$, $\mathcal T \subseteq\: \preccurlyeq$, and $\mathcal U \subseteq\: \preccurlyeq$.
\end{lemma}
Intuitively, by $\mathcal S$ a network-attached storage system may be simulated by a traditional storage system by forwarding public requests directed at $\textsc{nafs}$ to a hidden $\textsc{tfs}$ interface (via $\phi$). Symmetrically, by $\mathcal T$ a traditional storage system may be simulated by a network-attached storage system by forwarding public requests directed at $\textsc{tfs}$ to a hidden $\textsc{nafs}$ interface (via $\psi$). 
Finally, by $\mathcal U$ a network-attached storage system may simulate another network-attached storage system by filtering requests directed at $\textsc{nafs}$ through a hidden $\textsc{tfs}$ interface before forwarding them to a hidden $\textsc{nafs}$ interface (via $\phi[\psi]$). This rather mysterious detour forces a fresh capability to be acquired for every execution request.

By definition of $\lceil\_\rceil$ and alphaconversion to default public interfaces, we have for any $F$, $\rho$, and $C$:
\begin{enumerate}
\item $(\nu_{i \in \mathcal I}\alpha_i\beta_i)(C~|~\new{K_{M\!D}K'_M}\textsc{nafs}(F,\rho)) ~\preccurlyeq~ \phi[(\nu_{i \in \mathcal I}\betac_i)(\lceil C\rceil~|~\textsc{tfs}(\lceil F\rceil,\lceil \rho\rceil))] $
\item $(\nu_{i \in \mathcal I}\betac_i)(\lceil C\rceil~|~\textsc{tfs}(\lceil F\rceil,\lceil \rho\rceil)) ~\preccurlyeq ~ \psi[(\nu_{i \in \mathcal I}\alpha_i\beta_i)(C~|~\new{K_{M\!D}K'_M}\textsc{nafs}(F,\rho))] $
\item $\phi[\psi[(\nu_{i \in \mathcal I}\alpha_i\beta_i)(C~|~\new{K_{M\!D}K'_M}\textsc{nafs}(F,\rho))]] $
\par $~~~~~~~~~~~~~~~~~~~~~~~~~~~~~~~~~~~~~~~~~~~~~~~~~~~~~~~\preccurlyeq ~ (\nu_{i \in \mathcal I}\alpha_i\beta_i)(C~|~\new{K_{M\!D}K'_M}\textsc{nafs}(F,\rho))$
\end{enumerate}
Lemma \ref{static-lemma} follows by Proposition \ref{pf-prec}. Thus $\mathcal R$ is secure.

Further, Figure \ref{fig:simreln-4-s} shows a simulation relation for Lemma \ref{static-safetylemma}. We prove that the relation $\mathcal V$ is included in the simulation preorder.
\begin{lemma} $\mathcal V \subseteq \preccurlyeq$.
\end{lemma}
By definition of $\lceil\_\rceil$ and alphaconversion to default public interfaces, we have for any $F$, $\rho$, and $C$:
$$\psi[\phi[(\nu_{i \in \mathcal I}\betac_i)(\lceil C\rceil~|~\textsc{tfs}(\lceil F\rceil ,\lceil \rho\rceil))]]~\preccurlyeq ~ (\nu_{i \in \mathcal I}\beta_i)(\lceil C\rceil~|~\textsc{tfs}(\lceil F\rceil,\lceil \rho\rceil))$$
Lemma \ref{static-safetylemma} follows by Proposition \ref{pf-prec}. Thus $\mathcal R$ is safe and fully abstract.

\section{Models and proofs for dynamic access policies}\label{sec:mpdyn}
Next we present models and proofs for dynamic access policies, following the routine of Section \ref{sec:mpstat}.

\subsection{Models}
\begin{figure}
\hspace{-0.5cm}\fbox{\parbox{13.0cm}{\small
$$\infer[(Clk Req)]{k \in \mathbb N}{\textsc{tfs}(F,\Xi,\clk,\rho)^\eta
 \equiv \eta(\alphac_k)(x)\pause
\CReq(x)~|~\textsc{tfs}(F,\Xi,\clk,\rho)^\eta}
$$
$$\infer[(Time)]{k \in \mathbb N}{\CReq(M)~|~\textsc{tfs}(F,\Xi,\clk,\rho)^\eta
 \rightarrow \overline M\tup{\clk}~|~\textsc{tfs}(F,\Xi,\clk,\rho)^\eta}
$$
$$\infer[(Adm Req)]{k \in \mathbb N}{\textsc{tfs}(F,\Xi,\clk,\rho)^\eta
 \equiv \eta(\deltac_k)(\adm,x)\pause
\PReq_k(\adm,x)~|~\textsc{tfs}(F,\Xi,\clk,\rho)^\eta}
$$
$$\infer[(Adm Ok)]{\mathtt{perm}(F,k,\adm) = L \\ \func{push}(L,\adm,\Xi,\clk) = \tup{N,\Xi'}}
{\PReq_k(\adm,n)~|~\textsc{tfs}(F,\Xi,\clk,\rho)^\eta
\rightarrow  
\overline n\tup{N}~|~\textsc{tfs}(F,\Xi',\clk,\rho)^\eta}
$$
$$\infer[(Op Req)]{k \in \mathbb N}{\textsc{tfs}(F,\Xi,\clk,\rho)^\eta
 \equiv \eta(\betac_k)(\op,\tau,x)\pause
\Req_k(\op,\tau,x)~|~\textsc{tfs}(F,\Xi,\clk,\rho)^\eta}
$$
$$\infer[(Op Ok)]{\mathtt{perm}(F,k,\op) = L \\ \clk \leq \tau}
{\Req_k(\op,\tau,M)~|~\textsc{tfs}(F,\Xi,\clk,\rho)^\eta
\rightarrow  
\Comm(L,\op,M)~|~\textsc{tfs}(F,\Xi,\clk,\rho)^\eta}
$$
$$\infer[(Op Exec)]{\func{exec}(L,\op,\rho) = \tup{N,\rho'}}
{\Comm(L,\op,M)~|~\textsc{tfs}(F,\Xi,\clk,\rho)^\eta
\rightarrow 
\overline M\tup{N}~|~\textsc{tfs}(F,\Xi,\clk,\rho')^\eta}
%\qquad
%%
%\infer[(Op Res Ret)]{}{\Ret(n,r) \equiv }
$$
$$\infer[(Tick)]{\func{sync}(F,\Xi,\clk) = F'}
{\textsc{tfs}(F,\Xi,\clk,\rho)^\eta
\rightarrow 
\textsc{tfs}(F',\Xi,\clk+1,\rho)^\eta}
%\qquad
%%
%\infer[(Op Res Ret)]{}{\Ret(n,r) \equiv }
$$
%\line(1,0){370}
%
%$$\infer[(Dummy Clk Req)]{j \in \mathbb N\backslash \mathcal I}{{\shortuparrow^\mathrm{TS}_\mathrm{NAS}}^\eta \equiv \eta(\alpha_{j})(m)\pause \overline{\eta(\alphac_j)}\tup{m}~|~{\shortuparrow^\mathrm{TS}_\mathrm{NAS}}^\eta}
%\qquad
%
\fbox{\parbox{12.75cm}{
$$\infer[(Dummy Adm Req)]{j \in \mathbb N\backslash \mathcal I}{{\shortuparrow^\mathrm{TS}_\mathrm{NAS}}^\eta \equiv \eta(\delta_{j})(\op,x)\pause \overline{\eta(\deltac_j)}\tup{\op,x}~|~{\shortuparrow^\mathrm{TS}_\mathrm{NAS}}^\eta}$$
$$\infer[(Dummy Auth Req)]{j \in \mathbb N\backslash \mathcal I}{{\shortuparrow^\mathrm{TS}_\mathrm{NAS}}^\eta \equiv \eta(\alpha_{j})(\op,x)\pause 
\new m \overline{\eta(\alphac_j)}\tup{m}\pause m(\clk) \pause
 \overline x \tup{\func{mac}(\tup{j,\op,\clk},K_?}~|~{\shortuparrow^\mathrm{TS}_\mathrm{NAS}}^\eta}
%\qquad
%%
%\infer[(Dummy Auth Cap)]{}
%{\DAReq_{k}(\op,c) \equiv }
$$
$$\infer[(Dummy Exec Req)]{j \in \mathbb N\backslash \mathcal I}{{\shortuparrow^\mathrm{TS}_\mathrm{NAS}}^\eta \equiv \eta(\beta_{j})(\kappa,x) \pause\DReq(\kappa,x)^\eta~|~{\shortuparrow^\mathrm{TS}_\mathrm{NAS}}^\eta}
$$
$$\infer[(Dummy Op Req)]{\kappa = \func{mac}(\func{msg}(\kappa),K_?) \\
\func{msg}(\kappa) = \tup{\_,\op,\clk}}
{\DReq(\kappa,M)^\eta \rightarrow \overline{\eta(\betac_j)} \tup{\op,\clk,M}}$$
}}
}}
\caption{A traditional file system with local access control}
\label{fig:ts-sem-d}
\vspace{-0.2cm}
\end{figure}

\begin{figure}
\hspace{-0.5cm}\fbox{\parbox{13.0cm}{\small
%
%$$\infer[(Clk Req)]{k \in \mathbb N}{\textsc{nafs}(F,\Xi,\clk,\rho)^\eta
% \equiv \eta(\alpha_k)(m)\pause
%\CReq(m)~|~\textsc{nafs}(F,\Xi,\clk,\rho)^\eta}
%$$
%%
%$$\infer[(Time)]{k \in \mathbb N}{\CReq(m)~|~\textsc{nafs}(F,\Xi,\clk,\rho)^\eta
% \rightarrow \overline m\tup{\clk}~|~\textsc{nafs}(F,\Xi,\clk,\rho)^\eta}
%$$
%%
$$\infer[(Adm Req)]{k \in \mathbb N}{\textsc{nafs}(F,\Xi,\clk,\rho)^\eta
 \equiv \eta(\delta_k)(\adm,x)\pause
\PReq_k(\adm,x)~|~\textsc{nafs}(F,\Xi,\clk,\rho)^\eta}
$$
$$\infer[(Adm Ok)]{\mathtt{perm}(F,k,\adm) = L \\ \func{push}(L,\adm,\Xi,\clk) = \tup{N,\Xi'}}
{\PReq_k(\adm,M)~|~\textsc{nafs}(F,\Xi,\clk,\rho)^\eta
\rightarrow  
\overline M\tup{N}~|~\textsc{nafs}(F,\Xi',\clk,\rho)^\eta}
$$
$$\infer[(Auth Req)]{k \in \mathbb N}{\textsc{nafs}(F,\Xi,\clk,\rho)^\eta \equiv \eta(\alpha_{k})(\op,x)\pause \AReq_{k}(\op,x)~|~\textsc{nafs}(F,\Xi,\clk,\rho)^\eta}
$$
$$\infer[(Auth Cap)]{\mathtt{cert}(F,k,\op,\clk) = \kappa}
{\AReq_{k}(\op,M)~|~\textsc{nafs}(F,\Xi,\clk,\rho)^\eta \rightarrow \overline M\tup{\kappa}~|~\textsc{nafs}(F,\Xi,\clk,\rho)^\eta}
$$
%
%$$\infer[(Auth Bad Cap)]{\func{auth}(k,\op) \neq \func{ok}}
%{\AReq_{k}(\op,c)^\eta~|~\textsc{nafs}(F,\Xi,\clk,\rho)^\eta \rightarrow \overline c\tup {\func{mac}(\tup{k,\op},\mathrm K_\bot)}~|~\textsc{nafs}(F,\Xi,\clk,\rho)^\eta}$$
%
$$\infer[(Exec Req)]{k \in \mathbb N}{\textsc{nafs}(F,\Xi,\clk,\rho)^\eta \equiv \eta(\beta_{k})(\kappa,x)\pause \Req(\kappa,x)~|~\textsc{nafs}(F,\Xi,\clk,\rho)^\eta}
$$
$$\infer[(Op Ok)]{\mathtt{verif}(\kappa) = L \\ L \in \{\func{true},\func{false}\} \\ \func{msg}(\kappa) = \tup{\_,\op,\clk}}
{\Req(\kappa,M)~|~\textsc{nafs}(F,\Xi,\clk,\rho)^\eta \rightarrow  \Comm(L,\op,M)~|~\textsc{nafs}(F,\Xi,\clk,\rho)^\eta}$$
%
%$$\infer[(Exec Deny)]{\kappa \neq \func{mac}(\func{msg}(\kappa),\mathrm K)}
%{\Req(\kappa,n) \rightarrow 0}$$
%%
$$\infer[(Op Exec)]{\func{exec}(L,\op,\rho) = \tup{N,\rho'}}
{\Comm(L,\op,M)~|~\textsc{nafs}(F,\Xi,\clk,\rho)^\eta
\rightarrow 
\overline M\tup{N}~|~\textsc{nafs}(F,\Xi,\clk,\rho')^\eta}
%\qquad
%%
%\infer[(Op Res Ret)]{}{\Ret(n,r)
%\action{\new x\overline n\tup x} \{r/x\}}
$$
$$\infer[(Tick)]{\func{sync}(F,\Xi,\clk) = F'}
{\textsc{nafs}(F,\Xi,\clk,\rho)^\eta
\rightarrow 
\textsc{nafs}(F',\Xi,\clk+1,\rho)^\eta}
%\qquad
%%
%\infer[(Op Res Ret)]{}{\Ret(n,r) \equiv }
$$
%
%\line(1,0){370}
\fbox{\parbox{12.75cm}{
$$\infer[(Dummy Clk Req)]{j \in \mathbb N\backslash \mathcal I}{{\shortuparrow^\mathrm{NAS}_\mathrm{TS}}^\eta \equiv \eta(\alphac_{j})(x)\pause \new c\overline{\eta(\alpha_j)}\tup{x,c}\pause c(y)\pause \overline x\tup{\func{msg}(y).3}~|~{\shortuparrow^\mathrm{NAS}_\mathrm{TS}}^\eta}
$$
$$\infer[(Dummy Adm Req)]{j \in \mathbb N\backslash \mathcal I}{{\shortuparrow^\mathrm{NAS}_\mathrm{TS}}^\eta \equiv \eta(\deltac_{j})(\op,x)\pause \overline{\eta(\delta_j)}\tup{\op,x}~|~{\shortuparrow^\mathrm{NAS}_\mathrm{TS}}^\eta}$$
$$\infer[(Dummy Op Req)]{j \in \mathbb N\backslash \mathcal I}{{\shortuparrow^\mathrm{NAS}_\mathrm{TS}}^\eta \equiv \eta(\betac_{j})(\op,\tau,x)\pause \new c\overline{\eta(\alpha_{j})}\tup{\op,c} \pause c(\kappa)\pause [\func{msg}(\kappa).3 \leq \tau]\:\overline{\eta(\beta_j)}\tup{\kappa,x}~|~{\shortuparrow^\mathrm{NAS}_\mathrm{TS}}^\eta}
%\qquad
%\infer[(Dummy Exec Req)]{}{\DReq_j(\op,T,n)^\eta \equiv }
$$
}}
}}
\caption{A network-attached file system with distributed access control}
   \label{fig:nas-sem-d}
%\vspace{-0.3cm}
\end{figure}

The models extend those in Section \ref{sec:mpstat}, and are shown in Figures \ref{fig:ts-sem-d} and \ref{fig:nas-sem-d}. (As usual, we ignore the rules in the inner boxes in a first reading.) Interfaces are extended with channels $\delta_k$ and $\deltac_k$ for every $k$, on which users identified by $k$ send administration requests in the implementation and the specification. 

In the equational theory $\func{auth}(F,k,\op) = \func{ok}$ and $\func{exec}(L,\op,\rho) = \tup{N,\rho'}$ have the same meanings as in Section \ref{sec:mpstat}. Capabilities are derived by $\mathtt{cert}(\_,\_,\_,\_)$ as follows. 
%$$\infer
%	{L = \func{true}\mbox{ if }\func{auth}(F,k,\op) = \func{ok}\mbox{, }=\func{false}\mbox{ otherwise}}
%	{\mathtt{perm}(F,k,\op) \:\triangleq\: L}
%$$
$$\infer
	{a = K_{M\!D}\mbox{ if }\func{auth}(F,k,\op) = \func{ok}\mbox{, }=K'_M\mbox{ otherwise}}
	{\mathtt{cert}(F,k,\op,\clk) \:\triangleq\: \func{mac}(\tup{k,\op,\clk},a)}
$$
Recall that administrative operations scheduled at time $\clk$ are executed at the next clock tick (to $\clk+1$). In the equational theory $\func{push}(L,\adm,\Xi,\clk) = \tup{N,\Xi'}$ means that an administrative operation $\adm$ pushed on schedule $\Xi$ under decision $L$ at $\clk$ returns $N$ and the schedule $\Xi'$; and $\func{sync}(F,\Xi,\clk) = F'$ means that an access policy $F$ synchronized under schedule $\Xi$ at $\clk$ returns the access policy $F'$.

A traditional storage system may be described as
$$(\nu_{i \in \mathcal I}\alphac_i\betac_i\deltac_i)(C~|~\textsc{tfs}(F,\varnothing,0,\rho))$$
where $C$ is code run by honest users, $F$ is an access policy and $\rho$ is a store; initially the schedule is empty and the time is $0$.

Similarly a network-attached storage system may be described as
$$(\nu_{i \in \mathcal I}\alpha_i\beta_i\deltac_i)(C~|~\new{K_{M\!D}K'_M}\textsc{nafs}(F,\varnothing,0,\rho))$$

\begin{figure}
%\hspace{-1.0cm}
\hspace{-0.5cm}\fbox{\parbox{13.0cm}{\small
%\vspace{-0.3cm}
%$$\forall P,A.~~(\exists Q.~~\lceil P\rceil_A = Q)~ \Rightarrow ~$$ %\cup \{\betac_k~|~k \in \mathbb N\}$$
$$
\infer{\fn(M) \cap (\mathcal A \cup \{\alpha_j,\beta_j,\delta_j~|~j \in \mathbb N \backslash \mathcal I\}) = \varnothing}{\lceil M\rceil = M} \qquad
\infer{\lceil P\rceil_\Gamma = Q \\ \Gamma \supseteq \{\alpha_j,\beta_j,\delta_j~|~j \in \mathbb N \backslash \mathcal I\}}{\lceil P\rceil = Q}$$
$$\dots$$
$$
%  \infer{i \in \mathcal I \\ \fnv(m) \cap \dom(\Gamma) = \varnothing}{\lceil \overline{\alpha_i}\tup{m}\pause P\rceil_\Gamma = \overline{\alphac_i}\tup{m}\pause \lceil P\rceil_\Gamma}
%  \qquad
%
  \infer{i \in \mathcal I \\ \fnv(\adm,M) \cap \dom(\Gamma) = \varnothing}{\lceil \overline{\delta_i}\tup{\adm,M}\pause P\rceil_\Gamma = \overline{\deltac_i}\tup{\adm,M}\pause \lceil P\rceil_\Gamma}$$
$$
\infer{i \in \mathcal I \\ \fnv(c,x) \cap \dom(\Gamma) = \varnothing \\ c \notin \fn(P)}{\lceil \new c\overline{\alpha_i}\tup{\op,c}\pause c(x)\pause P\rceil_\Gamma   =
  \new c\overline{\alphac_i}\tup{c}\pause c(x)\pause \lceil P\rceil_{\Gamma, x : \mathtt{Cert}(i,\op)}} $$
%$$
%\infer{i \in \mathcal I \\ \fnv(c,x) \cap \dom(\Gamma) = \varnothing \\ \{c,x\} \cap \fnv(P) = \varnothing}{\lceil \new c\overline{\alpha_i}\tup{\op,c}\pause c(x)\pause \overline{\beta_i}\tup{x,M}\pause P\rceil_\Gamma   =
%  \overline{\betac_{i'}}\tup{\op,\infty,M}\pause \lceil P\rceil_{\Gamma, x : \mathtt{Cert}(i,\op)}} $$
$$
  \infer{\{i,i'\} \subseteq \mathcal I \\ \Gamma(x) = \mathtt{Cert}(i',\op) \\ \fnv(\op,M) \cap \dom(\Gamma) = \varnothing}{\lceil \overline{\beta_i}\tup{x,M}\pause P\rceil_\Gamma = \overline{\betac_{i'}}\tup{\op,x,M}\pause \lceil P\rceil_\Gamma}$$
}}
\caption{Abstraction function}
   \label{fig:abs-d}
%\vspace{-0.4cm}
\end{figure}

As usual, let $F$, $\rho$, and $C$ range over access policies, stores, and code for honest users that are ``wellformed" in the implementation, and let $\lceil\_\rceil$ abstract such $F$, $\rho$, and $C$ in the specification. We define
\begin{eqnarray*}
\mathcal R & = ~~~\bigcup_{F,\rho,C}\{ & (\nu_{i \in \mathcal I}\alphac_i\betac_i\deltac_i)(\lceil C\rceil~|~\textsc{tfs}(\lceil F\rceil,\varnothing,0,\lceil\rho\rceil))\\
&& (\nu_{i \in \mathcal I}\alpha_i\beta_i\delta_i)(C~|~\new{K_{M\!D}K'_M}\textsc{nafs}(F,\varnothing,0,\rho))~~~\}
\end{eqnarray*}
Figure \ref{fig:abs-d} shows the abstraction function $\lceil \_ \rceil$. Here
$$\mathcal A = \{{\alpha_j}_?, {\beta_j}_?,{\delta_j}_?,{\alphac_j}_?,{\betac_j}_?,{\deltac_j}_?~|~j \in \mathbb N\backslash \mathcal I\} \cup \{K_{M\!D},K'_M,K_?\} \cup \{\alpha_i,\beta_i,\delta_i~|~i \in \mathcal I\}$$

\subsection{Examples of security}
At this point we revisit the ``counterexamples" in Section \ref{intro-dyn}. By modeling them formally in this setting, we show that those counterexamples are eliminated. 

Recall (t1) and (t2).
\begin{description}
\item[t1] $\mathtt{acquire}~\kappa; \mathtt{chmod}~\zeta; \mathtt{use}~\kappa; \mathtt{success}\:\kappa$
\item[t2] $\mathtt{chmod}~\zeta; \mathtt{acquire}~\kappa; \mathtt{use}~\kappa; \mathtt{success}\:\kappa$
\end{description}
The following fragments of ${\it NAS}^d$ code formalize these traces.
\begin{description}
\item[I1] $(\nu c)~\overline{\alpha_i}\tup{\op,c}\pause c(\kappa)\pause (\nu m)~\overline{\delta_i}\tup{\zeta,m}\pause m(z)\pause (\nu n)~\overline{\beta_i}\tup{\kappa,n}\pause n(x)\pause [\mathtt{success}(x)]~\overline w\tup{}$
\item[I2] $(\nu m)~\overline{\delta_i}\tup{\zeta,m}\pause m(z)\pause (\nu c)~\overline{\alpha_i}\tup{\op,c}\pause c(\kappa)\pause (\nu n)~\overline{\beta_i}\tup{\kappa,n}\pause n(x)\pause [\mathtt{success}(x)]~\overline w\tup{}$
\end{description}
This code is abstracted to the following fragments of ${\it TS}^d$ code.
\begin{description}
\item[S1] $(\nu c)~\overline{\alphac_i}\tup{c}\pause c(\tau)\pause (\nu m)~\overline{\deltac_i}\tup{\zeta,m}\pause m(z)\pause (\nu n)~\overline{\betac_i}\tup{\op,\tau,n}\pause n(x)\pause [\mathtt{success}(x)]~\overline w\tup{}$
\item[S2] $(\nu m)~\overline{\deltac_i}\tup{\zeta,m}\pause m(z)\pause (\nu c)~\overline{\alphac_i}\tup{c}\pause c(\tau)\pause (\nu n)~\overline{\betac_i}\tup{\op,\tau,n}\pause n(x)\pause [\mathtt{success}(x)]~\overline w\tup{}$
\end{description}
Now whenever (I1) and (I2) can be distinguished, so can (S1) and (S2). Indeed the time bound $\tau$ is the same as the timestamp in $\kappa$; so (in particular) the operation request in (S1) is dropped whenever the execution request in (T1) is dropped.

A similar argument counters the ``dangerous" example with (t4) and (t5):
\begin{description}
\item[t4] $\mathtt{acquire}~\kappa; \mathtt{chmod}~\zeta'; \mathtt{acquire}~\kappa'; \mathtt{use}~\kappa'; \mathtt{success}\:\kappa'; \mathtt{use}~\kappa; \mathtt{success}\:\kappa$
\item[t5] $\mathtt{chmod}~\zeta'; \mathtt{acquire}~\kappa'; \mathtt{use}~\kappa'; \mathtt{success}\:\kappa'; \mathtt{acquire}~\kappa; \mathtt{use}~\kappa; \mathtt{success}\:\kappa$
\end{description}

Finally, recall (t8) and (t9).
\begin{description}
\item[t8] $\mathtt{acquire}\:\kappa; \mathtt{use}\:\kappa; c(); \mathtt{chmod}\:\zeta; c(); \mathtt{success}\:\kappa; \overline w\tup{}$
\item[t9] $c(); c(); \overline w\tup{}$
\end{description}
The following fragment of ${\it NAS}^d$ code formalizes (t8).
\begin{description}
\item[I3] $(\nu m)~\overline{\alpha_i}\tup{\op,m}\pause m(\kappa)\pause (\nu n)~\overline{\beta_i}\tup{\kappa,n}\pause $
\par $~~~~~~~~~~~~~~~~~~~~~~~~~~~~~~~~~~~~~~~~~~c() \pause (\nu m)~\overline{\delta_i}\tup{\zeta,m}\pause m(z)\pause c()\pause n(x)\pause [\mathtt{success}(x)]~\overline w\tup{}$
\end{description}
This code is abstracted to the following fragment of ${\it TS}^d$ code.
\begin{description}
\item[S3] $(\nu m)~\overline{\alphac_i}\tup{m}\pause m(\tau)\pause (\nu n)~\overline{\betac_i}\tup{\op,\tau,n}\pause $
\par $~~~~~~~~~~~~~~~~~~~~~~~~~~~~~~~~~~~~~~~~~~c() \pause (\nu m)~\overline{\deltac_i}\tup{\zeta,m}\pause m(z)\pause c()\pause n(x)\pause [\mathtt{success}(x)]~\overline w\tup{}$
\end{description}
A ${\it NAS}^d$ context distinguishes (I3) and (t9):
\begin{description}
\item $\overline c\tup{}; \overline{\alpha_j}\tup{\op',m_0}\pause m_0(\kappa'_0)\pause \overline{\beta_j}\tup{\kappa'_0,n_0}\pause n_0(x)\pause [\mathtt{failure}(x)]$
\par $~~~~~~~~~~~~~~~~~~~~\overline{\delta_j}\tup{\zeta,p}\pause  \overline{\alpha_j}\tup{\op',m_1}\pause m_1(\kappa'_1)\pause \overline{\beta_j}\tup{\kappa'_1,n_1}\pause n_1(x)\pause [\mathtt{success}(x)]~\overline c\tup{}$
\end{description}
But likewise a ${\it TS}^d$ context distinguishes (S3) and (t9):
\begin{description}
\item $\overline c\tup{}; \overline{\alpha_j}\tup{m_0}\pause m_0(\tau'_0)\pause \overline{\betac_j}\tup{\op',\tau'_0,n_0}\pause n_0(x)\pause [\mathtt{failure}(x)]$
\par $~~~~~~~~~~~~~~~~~~~~\overline{\deltac_j}\tup{\zeta,p}\pause  \overline{\alphac_j}\tup{m_1}\pause m_1(\tau'_1)\pause \overline{\betac_j}\tup{\op',\tau'_1,n_1}\pause n_1(x)\pause [\mathtt{success}(x)]~\overline c\tup{}$
\end{description}

\begin{figure}
\hspace{-0.5cm}\fbox{\parbox{13.0cm}{\small
$$\infer
	{\fn(\kappa,M) \cap \mathcal A = \varnothing \\ F',\clk' \leadsto F,\clk}
	{\Req(\kappa,M) ~~\mathcal S'^{F,\clk}_1~~ \DReq(\kappa,M)^{\eta_2}}
$$
$$\infer
	{k \in \mathbb N \\ \fn(\op,M) \cap \mathcal A = \varnothing \\ F',\clk' \leadsto F,\clk}
	{\Req(\mathtt{cert}(F',k,\op,\clk'),M) ~~\mathcal S'^{F,\clk}_1~~ \Req_k(\op,\clk',M)}
$$
$$\infer
	{\fn(L,\op,M) \cap \mathcal A = \varnothing}
	{\Comm(L,\op,M) ~~\mathcal S'^{F,\clk}_1~~ \Comm(L,\op,M)}
\qquad
\infer
	{k \in \mathbb N \\ \fn(\adm,M) \cap \mathcal A = \varnothing}
	{\PReq_k(\adm,M) ~~{\mathcal S'_1}^{F,\clk}~~ \PReq_k(\adm,M)}
$$
$$\infer
	{j \in \mathbb N\backslash \mathcal I \\ \fn(\op,M) \cap \mathcal A = \varnothing}
	{\AReq_j(\op,M) ~~\mathcal S'^{F,\clk}_1~~ \new m \overline{{\alphac_j}_?}\tup m\pause m(x)\pause \overline M\tup{\func{mac}(\tup{j,\op,x},\mathrm K_?)}}
$$
$$\inferc{file systems}
	{\forall r \in \mathcal L.~~P_r~~\mathcal S'^{F,\clk}_1~~Q_r \\ \fn(\Xi,\rho) \cap \mathcal A = \varnothing}
	{\textsc{nafs}(F,\Xi,\clk,\rho)~|~\Pi_{r \in \mathcal L} P_r ~~\mathcal S^{F,\clk}_1~~ \textsc{tfs}(F,\Xi,\clk,\rho)^{\eta_2}~|~\Pi_{r \in \mathcal L} Q_r}
$$
%$$\inferc{tfs-nafs}
%	{\forall x.~~x \in \mathtt{dom}(\sigma)~\Rightarrow~\exists F',\clk',i,\op.~~(F',\clk' \leadsto F,\clk) \wedge~\Gamma(x) = \mathtt{Cert}(i,\op) ~\wedge ~\sigma(x) = \mathtt{cert}(F',i,\op,\clk')}
%	{C\sigma ~~\mathcal S^{F,\clk}_2~~ \eta_3(\lceil C\rceil_\Gamma)}
%$$
$$\inferc{honest users}
	{\dom(\sigma) = \dom(\sigma') = X \\\\
	\!\!\!\!\!\!\!\!\!\!\!\!\!\!\!\!\!\!\!\!\!\!\!\!\!\!\!\!\!\!\!\!\!\!\!\!\!\!\!\forall x.~~x \in X~\Rightarrow~\exists F',\clk',i\in \mathcal I,\op.~~(F',\clk' \leadsto F,\clk)~\wedge~\sigma'(x) = \clk'~\\\\
	~~~~~~~~~~~~~~~~~~~~~~~~~~~~~~~~~~~~~~~~~~~~~~~~~~~~~~~~~~~ \wedge~\Gamma(x) = \mathtt{Cert}(i,\op) ~\wedge ~\sigma(x) = \mathtt{cert}(F',i,\op,\clk')}
	%\forall x.~~x \in \dom(\sigma'')~\Rightarrow~x \notin \dom(\Gamma) ~\wedge~\exists F',\clk',j\in \mathbb N \backslash \mathcal I,\op.~~(F',\clk' \leadsto F,\clk) \wedge~\sigma''(x) = \mathtt{cert}(F',j,\op,\clk')}
	{C\sigma ~~\mathcal S^{\Gamma,F,\clk}_2~~ \lceil C\rceil_\Gamma\sigma'}
$$
%$$\inferc{tfs-nafs}
%	{\Gamma(x) = \mathtt{Cert}(i,\op)}
%	{(\nu c)(c(x)\pause C~|~\AReq_i(\op,c)) ~~\mathcal S'_3~~ \eta_3(\lceil C\rceil_\Gamma)}
%\qquad
%\inferc{tfs-nafs}
%	{P~~\mathcal S_2~~Q}
%	{(\nu c)(c(x)\pause P~|~\overline c\tup{\mathtt{cert}(i,\op)}) ~~\mathcal S'_3~~ Q}
$$\infer
	{i \in \mathcal I \\ P~~\mathcal S^{\Gamma,F,\clk}_2~~ Q \\ \Gamma(x) = \mathtt{Cert}(i,\op)}
	{(\nu c)(c(x)\pause P~|~\AReq_i(\op,c)) ~~\mathcal S_3^{F,\clk}~~ (\nu c)(c(x)\pause Q~|~\CReq(c))}
$$
$$\inferc{trusted code}
	{P~~\mathcal S^{F,\clk}_1~~Q \\ P'~~\mathcal S^{\Gamma,F,\clk}_2~~Q' \\ \forall r \in \mathcal L.~~P_r~~\mathcal S_3^{F,\clk}~~Q_r}
	{(\nu_{i \in \mathcal I}\alpha_i\beta_i\delta_i)(P~|~P'~|~\Pi_{r \in \mathcal L} P_r) ~~\mathcal S'^{F,\clk}~~ (\nu_{i \in \mathcal I}\alphac_i\betac_i\deltac_i)(Q~|~Q'~|~\Pi_{r \in \mathcal L} Q_r)}
$$
$$\inferc{system code}
	{P~~\mathcal S'^{F,\clk}~~Q \\ \forall x,N.~~(\exists \sigma'.~~\sigma \equiv \subs N x~|~\sigma')~\Rightarrow~N:_{\mathcal F,F,\clk} \mathtt{Export}}
	{(\nu \seq n)(\nu K_{M\!D}K'_M)(\sigma~|~P)~~\mathcal S^{F,\clk}~~(\nu \seq n)(\nu K_?)(\eta_3(\sigma)~|~(\nu_{j \in \mathbb N\backslash\mathcal I}{\alphac_j}_?{\betac_j}_?{\deltac_j}_?)(Q~|~{\shortuparrow^\mathrm{TS}_\mathrm{NAS}}))}
$$
%$$\inferc{tfs-nafs-Final-CUT}
%	{P~~\mathcal S'^{F,\clk}~~Q \\ \forall x,M.~~(\exists \sigma'.~~\sigma \equiv \subs M x~|~\sigma')~\Rightarrow~M: \mathtt{Export}}
%	{(\nu \seq n)(\nu \mathrm K\mathrm K_\bot)(\sigma~|~(\nu_{i \in \mathcal I}\alpha_i\beta_i)~P)~~\mathcal S~~(\nu \seq n)(\nu \mathrm K_?)(\eta_3(\sigma)~|~(\nu_{j \in \mathbb N\backslash\mathcal I}\betac_j)((\nu_{i \in \mathcal I}\betac_i)~Q~|~\bowtie\!\textsc I))}
%$$
}}
\caption{Simulation relation for Lemma \ref{dynamic-lemma}.1 ($\_ \preccurlyeq \phi[\_]$)}
\label{fig:simreln-1-d}
\end{figure}

\begin{figure}
\hspace{-0.5cm}\fbox{\parbox{13.0cm}{\small
$$\infer
	{i \in \mathcal I \\ \fn(\op,M) \cap \mathcal A = \varnothing \\ F',\clk' \leadsto F,\clk }
	{\Req_i(\op,\clk',M) ~~{\mathcal T'_1}^{F,\clk}~~ \Req(\mathtt{cert}(F',k,\op,\clk'),M)^{\eta_1}}
$$
$$\infer
	{j \in \mathbb N \backslash \mathcal I \\ \fn(\op,\tau,M) \cap \mathcal A = \varnothing}
	{\Req_j(\op,\tau,M) ~~{\mathcal T'_1}^{F,\clk}~~ \new c\overline{{\alpha_j}_?}\tup{\op,c} \pause c(\kappa)\pause [\func{msg}(\kappa).3 \leq \tau]\:\overline{{\beta_j}_?}\tup{\kappa,M}}$$
%$$\inferc{tfs-nafs}
%	{k \in \mathbb N \\ \fn(\op,M) \cap \mathcal A = \varnothing \\ F',\clk' \leadsto F,\clk}
%	{\Req_k(\op,T,n) ~~{\mathcal T'_1}^{F,\clk}~~ \Comm(\op,n)^{\eta_1}}
%\quad
$$\infer
	{\fn(L,\op,M) \cap \mathcal A = \varnothing}
	{\Comm(L,\op,M) ~~{\mathcal T'_1}^{F,\clk}~~ \Comm(L,\op,M)^{\eta_1}}
$$
$$
\infer
	{k \in \mathbb N \\ \fn(\adm,M) \cap \mathcal A = \varnothing}
	{\PReq_k(\adm,n) ~~{\mathcal T'_1}^{F,\clk}~~ \PReq_k(\adm,n)^{\eta_1}}
$$%
$$\infer
	{j \in \mathbb N \backslash \mathcal I \\ \fn(M) \cap \mathcal A = \varnothing}
	{\CReq(M) ~~{\mathcal T'_1}^{F,\clk}~~ \new c\overline{{\alpha_j}_?}\tup{M,c}\pause c(x)\pause \overline M\tup{\func{msg}(x).3}}
$$
$$\inferc{file systems}
	{\forall r \in \mathcal L.~~P_r~~{\mathcal T'_1}^{F,\clk}~~Q_r \\ \fn(\Xi,\rho) \cap \mathcal A = \varnothing}
	{\textsc{tfs}(F,\Xi,\clk,\rho)~|~\Pi_{r \in \mathcal L} P_r ~~{\mathcal T_1}^{F,\clk}~~ \textsc{nafs}(F,\Xi,\clk,\rho)^{\eta_1}~|~\Pi_{r \in \mathcal L} Q_r }
%\qquad
%%$$\inferc{tfs-nafs}
%%	{\forall x.~~x \in \mathtt{dom}(\sigma)~\Rightarrow~\exists i,\op.~~\Gamma(x) = \mathtt{Cert}(i,\op) ~\wedge~\sigma(x) = \mathtt{cert}(F,\clk,i,\op)}
%%	{ \lceil C\rceil_\Gamma~~{\mathcal T_2}^{F,\clk}~~ \mathcal T^\star(C)\sigma}
%%$$
%\inferc{tfs-nafs}
%	{}
%	{ \lceil C\rceil_\Gamma~~{\mathcal T_2}^{F,\clk}~~ \mathcal T^\star(C)}
$$
$$\inferc{honest users}
	{\dom(\sigma) = \dom(\sigma') = X \\\\ \forall x.~~x \in X~\Rightarrow~\exists F',\clk',i\in \mathcal I,\op.~~(F',\clk' \leadsto F,\clk)~\wedge~ \sigma(x) = \clk' ~~~~~~~~~~~~~~~~~~~~~~~~~~~~~~ \\\\
	~~~~~~~~~~~~~~~~~~~~~~~~~~~~~~~~~~~~~~~~~~~~~~~~~~~~~~~~~~\wedge~\Gamma(x) = \mathtt{Cert}(i,\op) ~\wedge~\sigma'(x) = \mathtt{cert}(F',i,\op,\clk')}
	{ \lceil C\rceil_\Gamma\sigma~~{\mathcal T_2}^{\Gamma,F,\clk}~~ C\sigma'}
$$
$$\infer
	{i \in \mathcal I \\ P~~{\mathcal T_2}^{\Gamma,F,\clk}~~ Q \\ \Gamma(x) = \mathtt{Cert}(i,\op)}
	{(\nu c)(c(x)\pause P~|~\CReq(c)) ~~\mathcal T'_3~~ (\nu c)(c(x)\pause Q~|~\AReq_i(\op,c))}
%\quad
%\inferc{tfs-nafs-ALT}
%	{\Gamma(x) = \mathsf{Cert}(i,\op) \\ \clk \leq \clk' \Rightarrow \clk' = \clk \wedge F' = F}
%	{(\nu c)(c(x)\pause \lceil C\rceil_\Gamma~|~\overline c\tup{\clk'}) ~~\mathcal T'^{F,\clk}_3~~ (\nu c)(c(x)\pause C~|~\overline c\tup{\mathtt{cert}(F',\clk',i,\op)})}
$$
$$\inferc{trusted code}
	{P~~\mathcal T^{F,\clk}_1~~Q \\ P'~~\mathcal T^{\Gamma,F,\clk}_2~~Q' \\ \forall r \in \mathcal L.~~P_r~~\mathcal T'_3~~Q_r}
	{(\nu_{i \in \mathcal I}\alphac_i\betac_i\deltac_i)(P~|~P'~|~\Pi_{r \in \mathcal L} P_r) ~~\mathcal T'~~ (\nu_{i \in \mathcal I}\alpha_i\beta_i\delta_i)(\nu K_{M\!D}K'_M)(Q~|~Q'~|~\Pi_{r \in \mathcal L} Q_r)}
$$
$$\inferc{system code}
	{P~~{\mathcal T'}~~Q}
	{(\nu \seq n)(\sigma~|~P) ~~\mathcal T~~ (\nu \seq n)(\sigma~|~(\nu_{j \in \mathbb N\backslash\mathcal I}{\alpha_j}_?{\beta_j}_?{\delta_j}_?)(Q~|~{\shortuparrow^\mathrm{NAS}_\mathrm{TS}}))}
$$
%$$\inferc{tfs-nafs-Final-CUT}
%	{P_1~~\mathcal T_1~~Q_1 \\ P_2~~\mathcal T_2~~Q_2}
%	{(\nu \seq n)(\sigma~|~(\nu_{i \in \mathcal I}\betac_i)(P_1~|~P_2)) ~~\mathcal T~~ (\nu \seq n)(\sigma~|~(\nu_{j \in \mathbb N\backslash\mathcal I}\alpha_j\beta_j)((\nu_{i \in \mathcal I}\alpha_i\beta_i)(\new{\mathrm K\mathrm K_\bot}Q_1~|~Q_2)~|~{\shortuparrow^\mathrm{NAS}_\mathrm{TS}}))}
%$$
}}
\caption{Simulation relation for Lemma \ref{dynamic-lemma}.2 ($\_ \preccurlyeq \psi[\_]$)}
\label{fig:simreln-2-d}
\end{figure}

%\begin{lemma} Suppose that $M:\mathtt{Export}$ and $N:\mathtt{Export}$. Then $M = N$ iff $\eta_3(M) = \eta_3(N)$.
%\end{lemma}

\begin{figure}
\vspace{-0.5cm}
\hspace{-0.5cm}\fbox{\parbox{13.0cm}{\small
$$\infer
	{\fn(\kappa,M) \cap \mathcal A = \varnothing}
	{\DReq(\kappa,M)^{\eta_2} ~~\mathcal U'^{F,\clk}_1~~ \Req(\kappa,M)}
\qquad
\infer
	{j \in \mathbb N \backslash \mathcal I \\ \fn(\op,M) \cap \mathcal A = \varnothing \\ F',\clk' \leadsto F,\clk}
	{\overline{{\betac_j}_?}\tup{\op,\clk',M} ~~\mathcal U'^{F,\clk}_1~~ \Req(\mathtt{cert}(F',j,\op,\clk') ,M)}
$$
$$\infer
	{j \in \mathbb N \backslash \mathcal I \\ \fn(\op,M) \cap \mathcal A = \varnothing \\ F',\clk' \leadsto F,\clk}
	{\DReq_j(\op,\clk',M)^{\eta_1\oplus\eta_2} ~~\mathcal U'^{F,\clk}_1~~ \Req(\mathtt{cert}(F',j,\op,\clk') ,M)}
$$
$$\infer
	{j \in \mathbb N \backslash \mathcal I \\ \fn(\op,M) \cap \mathcal A = \varnothing \\ F',\clk' \leadsto F,\clk}
	{(\nu c)(c(x)\pause [\func{msg}(x).3 \leq \clk']\:\overline{{\beta_j}_?}\tup{x,M}~|~\AReq_j(\op,c)) ~~\mathcal U'^{F,\clk}_1~~ \Req(\mathtt{cert}(F',j,\op,\clk'),M)}
$$
$$\infer
	{j \in \mathbb N \backslash \mathcal I \\ \fn(\op,M) \cap \mathcal A = \varnothing \\ F',\clk' \leadsto F,\clk \\ N = \func{mac}(\tup{j,\op,\clk'},K_?)}
	{(\nu c)(c(x)\pause [\func{msg}(x).3 \leq \clk']\:\overline{{\beta_j}_?}\tup{x,M}~|~\overline c\tup{N}) ~~\mathcal U'^{F,\clk}_1~~ \Req(\mathtt{cert}(F',j,\op,\clk'),M)}
$$
$$\infer
	{j \in \mathbb N \backslash \mathcal I \\ \fn(\op,M) \cap \mathcal A = \varnothing \\ F',\clk' \leadsto F,\clk}
	{\overline{{\beta_j}_?}\tup{\func{mac}(\tup{j,\op,\clk'},K_?),M} ~~\mathcal U'^{F,\clk}_1~~ \Req(\mathtt{cert}(F',j,\op,\clk') ,M)}
$$
$$\infer
	{k \in \mathbb N \\ \fn(\op,M) \cap \mathcal A = \varnothing \\ F',\clk' \leadsto F,\clk}
	{\Req(\func{mac}(\tup{k,\op,\clk'},K_?),M) ~~\mathcal U'^{F,\clk}_1~~ \Req(\mathtt{cert}(F',k,\op,\clk'),M)}
$$
$$\infer
	{\fn(L,\op,M) \cap \mathcal A = \varnothing}
	{\Comm(L,\op,M) ~~\mathcal U'^{F,\clk}_1~~ \Comm(L,\op,M)}
$$
$$\infer
	{j \in \mathbb N \backslash \mathcal I \\ \fn(\op,M) \cap \mathcal A = \varnothing}
	{\new m\overline{{\alphac_j}_?}\tup{m}\pause m(x)\pause  \overline M\tup{\func{mac}(\tup{j,\op,x},K_?)}~~\mathcal U'^{F,\clk}_1~~ \AReq_j(\op,M)}
$$
$$\infer
	{j \in \mathbb N \backslash \mathcal I \\ \fn(\op,M) \cap \mathcal A = \varnothing \\ F',\clk' \leadsto F,\clk}
	{\new m (m(x)\pause  \overline M\tup{\func{mac}(\tup{j,\op,x}, K_?)}~|~\overline m\tup{\clk'})~~\mathcal U'^{F,\clk}_1~~ \overline M\tup{\mathtt{cert}(F',k,\op,\clk')}}
$$
$$\infer
	{\fn(\adm,M) \cap \mathcal A = \varnothing}
	{\PReq_k(\adm,M) ~~\mathcal U'^{F,\clk}_1~~ \PReq_k(\adm,M)}
$$
%$$\inferc{tfs-nafs}
%	{}
%	{\overline c\tup{\func{mac}(\tup{\tup{j,\eta_3(\op)},\clk+1},\mathrm K_?)}~~\mathcal U'_1~~ \AReq_j(\op,c)}
%$$
$$\inferc{file systems}
	{\forall r \in \mathcal L.~~P_r~~\mathcal U'^{F,\clk}_1~~Q_r \\ \fn(\Xi,\rho) \cap \mathcal A = \varnothing}
	{{\shortuparrow^\mathrm{TS}_\mathrm{NAS}}^{\eta_2}~|~{\shortuparrow^\mathrm{NAS}_\mathrm{TS}}^{\eta_1\oplus\eta_2}~|~\textsc{nafs}(F,\Xi,\clk,\rho)^{\eta_1}~|~\Pi_{r \in \mathcal L} P_r ~~\mathcal U^{F,\clk}_1~~ \textsc{nafs}(F,\Xi,\clk,\rho)~|~\Pi_{r \in \mathcal L} Q_r}
$$
$$\inferc{honest users}
	{\lceil C\rceil_\Gamma = C^\circ \\\\ 
	\forall x.~~x \in \dom(\sigma)~\Rightarrow~\exists F',\clk',i \in \mathcal I,\op.~~(F',\clk' \leadsto F,\clk)~~~~~~~~~~~~~~~~~~~~~~~~~~~~~~~~~~~~~~~~~~~~~~~~~~~~\\\\
	~~~~~~~~~~~~~~~~~~~~~~~~~~~~~~~~~~~~~~~~~~~~~~~~~~~~~~~~~~~~~~~~~~~~~\wedge~\Gamma(x) = \mathtt{Cert}(i,\op) ~\wedge ~\sigma(x) = \mathtt{cert}(F',i,\op,\clk')}
	{C\sigma ~~\mathcal U^{\Gamma,F,\clk}_2~~ C\sigma}
$$
$$\infer
	{i \in \mathcal I \\ \Gamma(x) = \mathtt{Cert}(i,\op) \\ P~~\mathcal U^{\Gamma,F,\clk}_2~~Q}
	{(\nu c)(c(x)\pause P~|~\AReq_i(\op,c)) ~~\mathcal U^{F,\clk}_3~~ (\nu c)(c(x)\pause Q~|~\AReq_i(\op,c))}
$$
$$\inferc{trusted code}
	{P~~\mathcal U^{F,\clk}_1~~Q \\ P'~~\mathcal U^{F,\clk}_2~~Q' \\ \forall r \in \mathcal L.~~P_\ell~~\mathcal U^{F,\clk}_3~~Q_\ell}
	{(\nu_{i \in \mathcal I}\alpha_i\beta_i\delta_i)(\nu K_{M\!D} K'_M)(P~|~P'~|~\Pi_{r \in \mathcal L} P_r) ~~\mathcal U'^{F,\clk}~~ (\nu_{i \in \mathcal I}\alpha_i\beta_i\delta_i)(Q~|~Q'~|~\Pi_{r \in \mathcal L} Q_r)}
$$
%$$\inferc{tfs-nafs}
%	{P~~\mathcal U'~~Q \\ \forall x,M.~~(\exists \sigma'.~~\sigma \equiv \subs M x~|~\sigma')~\Rightarrow~M \notin \{\mathrm K,\mathrm K_\bot\}}
%	{(\nu \seq n)(\nu \mathrm K\mathrm K_\bot)(\sigma~|~(\nu_{i \in \mathcal I}\alpha_i\beta_i)~P)~~\mathcal U~~(\nu \seq n)(\nu \mathrm K_?)(\eta_3(\sigma)~|~(\nu_{j \in \mathbb N\backslash\mathcal I}\betac_j)(\nu\mathrm K\mathrm K_\bot)((\nu_{i \in \mathcal I}\betac_i)~Q~|~\bowtie\!\textsc I))}
%$$
$$\inferc{system code}
	{P~~\mathcal U'^{F,\clk}~~Q \\ \forall x,N.~~(\exists \sigma'.~~\sigma \equiv \subs N x~|~\sigma')~\Rightarrow~N:_{\mathcal F,F,\clk} \mathtt{Export}}
	{(\nu \seq n)(\nu K_?)(\eta_3(\sigma)~|~(\nu_{j \in\mathbb N\backslash\mathcal I}{\alphac_j}_?{\betac_j}_?{\deltac_j}_?{\alpha_j}_?{\beta_j}_?{\delta_j}_?)~P)~~\mathcal U~~(\nu \seq n)(\nu K_{M\!D}K'_M)(\sigma~|~Q)}
$$
}}
\caption{Simulation relation for Lemma \ref{dynamic-lemma}.3 ($\phi[\psi[\_]] \preccurlyeq \_$)}
\label{fig:simreln-3-d}
\end{figure}

\subsection{Proofs of security}

We show that $\mathcal R$ is secure, safe, and fully abstract. Recall the contexts $\phi$ and $\psi$ defined in Section \ref{sec:mpstat}. The processes $\shortuparrow^\mathrm{NAS}_\mathrm{TS}$ and $\shortuparrow^\mathrm{TS}_\mathrm{NAS}$ are redefined in the inner boxes in Figures \ref{fig:ts-sem-d} and \ref{fig:nas-sem-d}. In particular, the rule \textsc{(Dummy Op Req)} in Figure \ref{fig:nas-sem-d} translates time-bounded operation requests by ${\it TS}^d$ contexts. 

Simulation relations for security are shown in Figures \ref{fig:simreln-1-d}, \ref{fig:simreln-2-d}, and \ref{fig:simreln-3-d}, and a simulation relation for safety and full abstraction is shown in Figure \ref{fig:simreln-4-d}. Here
\begin{eqnarray*}
\eta_1 & \triangleq & [\alpha_j \mapsto {\alpha_j}_?,\beta_j \mapsto {\beta_j}_?,\delta_j \mapsto {\delta_j}_?~|~j \in \mathbb N\backslash \mathcal I]\\
\eta_2 & \triangleq & [\alphac_j \mapsto {\alphac_j}_?,\betac_j \mapsto {\betac_j}_?,\deltac_j \mapsto {\deltac_j}_?~|~j \in \mathbb N\backslash \mathcal I]
\end{eqnarray*}
A binary relation $\_,\_ \leadsto\_,\_$ (``leads-to") is defined over the product of access policies and clocks. Access policies may change at clock ticks (but not between).
$$F',\clk' \leadsto F,\clk \:\triangleq\: (\clk' < \clk) \vee (\clk' = \clk \wedge F' = F)$$
As usual, any term that may be available to contexts must be of type $\mathtt{Export}$.
$$\infer
	{N = N'\sigma \\ \{K_{M\!D},K'_M,K_?\}\cap \fn(N') = \varnothing \\ \forall L \in \mathtt{rng}(\sigma).~~\exists j \in \mathbb N \backslash \mathcal I,\op,\clk'.~~\op :_{\mathcal F,F,\clk} \mathtt{Export}~\wedge~(\mathcal F(\clk'),\clk' \leadsto F,\clk)\\\\
	~~~~~~~~~~~~~~~~~~~~~~~~~~\wedge~L = \mathtt{cert}(\mathcal F(\clk'),j,\op,\clk')}
	{N :_{\mathcal F,F,\clk} \mathtt{Export}}
$$
\begin{figure}
\vspace{-1.0cm}
\hspace{-1.8cm}\fbox{\parbox{15.6cm}{\small
$$\infer
	{j \in \mathbb N \backslash \mathcal I \\ \fn(\op,\tau,M) \cap \mathcal A = \varnothing}
	{\new c\overline{{\alpha_j}_?}\tup{\op,c} \pause c(\kappa)\pause [\func{msg}(\kappa).3 \leq \tau]\:\overline{{\beta_j}_?}\tup{\kappa,M} ~~\mathcal V'^{F,\clk}_1~~ \Req_j(\op,\tau,M)}
$$
$$\infer
	{j \in \mathbb N \backslash \mathcal I \\ \fn(\op,\tau,M) \cap \mathcal A = \varnothing \\ N = \func{mac}(\tup{j,\op,x},K_?)}
	{(\nu c)(c(\kappa)\pause [\func{msg}(\kappa).3 \leq \tau]\:\overline{{\beta_j}_?}\tup{\kappa,M}~|~\new m\overline{{\alphac_j}_?}\tup{m}\pause m(x)\pause  \overline c\tup{N}) ~~\mathcal V'^{F,\clk}_1~~ \Req_j(\op,\tau,M)}
$$
$$\infer
	{j \in \mathbb N \backslash \mathcal I \\ \fn(\op,\tau,M) \cap \mathcal A = \varnothing \\ N = \func{mac}(\tup{j,\op,x},K_?)}
	{(\nu c)(c(\kappa)\pause [\func{msg}(\kappa).3 \leq \tau]\:\overline{{\beta_j}_?}\tup{\kappa,M}~|~(\nu m)(m(x)\pause  \overline c\tup{N}~|~\CReq(m))) ~~\mathcal V'^{F,\clk}_1~~ \Req_j(\op,\tau,M)}
$$
$$\infer
	{j \in \mathbb N \backslash \mathcal I \!\!\\\!\! \fn(\op,\tau,M) \cap \mathcal A = \varnothing \\ F',\clk' \leadsto F,\clk \\ N = \func{mac}(\tup{j,\op,\clk'},K_?) \\ L = \mathtt{perm}(F',j,\op)}
	{(\nu c)(c(\kappa)\pause [\func{msg}(\kappa).3 \leq \tau]\:\overline{{\beta_j}_?}\tup{\kappa,M}~|~(\nu m)(m(x)\pause  \overline c\tup{N}~|~\overline m\tup{\clk'})) ~~\mathcal V'^{F,\clk}_1~~ [\clk \leq \tau]~\Comm(L,\op,M)}
$$
$$\infer
	{j \in \mathbb N \backslash \mathcal I \\ \fn(\op,\tau,M) \cap \mathcal A = \varnothing \\ F',\clk' \leadsto F, \clk  \\ L = \mathtt{perm}(F',j,\op)}
	{(\nu c)(c(\kappa)\pause [\func{msg}(\kappa).3 \leq \tau]\:\overline{{\beta_j}_?}\tup{\kappa,M}~|~\overline c\tup{\func{mac}(\tup{j,\op,\clk'},K_?)}) ~~\mathcal V'^{F,\clk}_1~~ [\clk \leq \tau]~\Comm(L,\op,M)}
$$
$$\infer
	{j \in \mathbb N \backslash \mathcal I \\ \fn(\op,M) \cap \mathcal A = \varnothing \\ F',\clk' \leadsto F,\clk \\\\
	N = \func{mac}(\tup{j,\op,\clk'},K_?)  \\ L = \mathtt{perm}(F',j,\op)}
	{\overline{{\beta_j}_?}\tup{N,M} ~~\mathcal V'^{F,\clk}_1~~ \Comm(L,\op,M)}
\quad~~
\infer
	{j \in \mathbb N \backslash \mathcal I \\ \fn(\op,M) \cap \mathcal A = \varnothing \\ F',\clk' \leadsto F,\clk \\\\
	N = \func{mac}(\tup{j,\op,\clk'},K_?) \\ L = \mathtt{perm}(F',j,\op)}
	{\DReq(N,M)^{\eta_1\oplus\eta_2} ~~\mathcal V'^{F,\clk}_1~~ \Comm(L,\op,M)}
$$
$$\infer
	{j \in \mathbb N \backslash \mathcal I \\ \fn(\op,M) \cap \mathcal A = \varnothing \\\\ F',\clk' \leadsto F,\clk  \\ L = \mathtt{perm}(F',j,\op)}
	{\overline{{\betac_j}_?}\tup{\op,\clk',M} ~~\mathcal V'^{F,\clk}_1~~ \Comm(L,\op,M)}
\qquad
\infer
	{\fn(\op,M) \cap \mathcal A = \varnothing}
	{\Comm(L,\op,M) ~~\mathcal V'^{F,\clk}_1~~ \Comm(L,\op,M)}
$$
$$\infer
	{\fn(\adm,M) \cap \mathcal A = \varnothing}
	{\PReq_k(\adm,M) ~~\mathcal V'^{F,\clk}_1~~ \PReq_k(\adm,M)}
\qquad
\infer
	{j \in \mathbb N \backslash \mathcal I \\ \fn(M) \cap \mathcal A = \varnothing}
	{\new c\overline{{\alpha_j}_?}\tup{M,c}\pause c(y)\pause \overline M\tup{\func{msg}(y).3}~~\mathcal V'^{F,\clk}_1~~ \CReq(M)}
$$
$$\infer
	{j \in \mathbb N \backslash \mathcal I \\ \fn(M) \cap \mathcal A = \varnothing}
	{(\nu c)(c(y)\pause \overline M\tup{\func{msg}(y).3}~|~(\nu m)~\overline{{\alphac_j}_?}\tup{m}\pause m(x)\pause  \overline c\tup{\func{mac}(\tup{j,M,x},K_?)})
~~\mathcal V'^{F,\clk}_1~~ \CReq(M)}
$$
$$\infer
	{j \in \mathbb N \backslash \mathcal I \\ \fn(M) \cap \mathcal A = \varnothing}
	{(\nu c)(c(y)\pause \overline M\tup{\func{msg}(y).3}~|~(\nu m)(m(x)\pause  \overline c\tup{\func{mac}(\tup{j,M,x},K_?)}
~|~\CReq(M)))~~\mathcal V'^{F,\clk}_1~~ \CReq(M)}
$$
$$\infer
	{j \in \mathbb N \backslash \mathcal I \\ \fn(M) \cap \mathcal A = \varnothing \\ \clk' \leq \clk}
	{(\nu c)(c(y)\pause \overline M\tup{\func{msg}(y).3}~|~(\nu m)(m(x)\pause  \overline c\tup{\func{mac}(\tup{j,M,x},K_?)}
~|~\overline m\tup{\clk'}))~~\mathcal V'^{F,\clk}_1~~ \overline m\tup{\clk'}}
$$
$$\infer
	{j \in \mathbb N \backslash \mathcal I \\ \fn(M) \cap \mathcal A = \varnothing}
	{(\nu c)(c(y)\pause \overline M\tup{\func{msg}(y).3}~|~\overline c\tup{\func{mac}(\tup{j,M,\clk'},K_?)})~~\mathcal V'^{F,\clk}_1~~ \overline M\tup{\clk'}}
$$
%$$\inferc{tfs-nafs}
%	{}
%	{\overline c\tup{\func{mac}(\tup{\tup{j,\eta_3(\op)},\clk+1},\mathrm K_?)}~~\mathcal V'_1~~ \AReq_j(\op,c)}
%$$
$$\inferc{file systems}
	{\forall r \in \mathcal L.~~P_r~~\mathcal V'^{F,\clk}_1~~Q_r \\ \fn(\Xi,\rho) \cap \mathcal A = \varnothing}
	{{\shortuparrow^\mathrm{NAS}_\mathrm{TS}}^{\eta_1}~|~{\shortuparrow^\mathrm{TS}_\mathrm{NAS}}^{\eta_1\oplus\eta_2}~|~\textsc{tfs}(F,\Xi,\clk,\rho)^{\eta_2}~|~\Pi_{r \in \mathcal L} P_r ~~\mathcal V^{F,\clk}_1~~ \textsc{tfs}(F,\Xi,\clk,\rho)~|~\Pi_{r \in \mathcal L} Q_r}
$$
$$\inferc{honest users}
	{\forall x.~~x \in \dom(\sigma)~\Rightarrow~\exists \clk',i \in \mathcal I,\op.~~\clk' \leq \clk~\wedge~\Gamma(x) = \mathtt{Cert}(i,\op) ~\wedge ~\sigma(x) = \clk'}
	{\lceil C\rceil_\Gamma\sigma ~~\mathcal V^{\Gamma,F,\clk}_2~~ \lceil C\rceil_\Gamma\sigma}
$$
$$\infer
	{i \in \mathcal I \\ \Gamma(x) = \mathtt{Cert}(i,\op) \\ P~~\mathcal V^{\Gamma,F,\clk}_2~~Q}
	{(\nu c)(c(x)\pause P~|~\CReq(c)) ~~\mathcal V^{F,\clk}_3~~ (\nu c)(c(x)\pause Q~|~\CReq(c))}
$$
$$\inferc{system code}
	{P~~\mathcal V^{F,\clk}_1~~Q \\ P'~~\mathcal V^{F,\clk}_2~~Q' \\ \forall r \in \mathcal L.~~P_\ell~~\mathcal V^{F,\clk}_3~~Q_\ell \\\\
	P'' = (\nu_{i \in \mathcal I}\alphac_i\betac_i\deltac_i)(\nu K_?)(P~|~P'~|~\Pi_{r \in \mathcal L} P_r) \\ Q'' = (\nu_{i \in \mathcal I}\alphac_i\betac_i\deltac_i)(Q~|~Q'~|~\Pi_{r \in \mathcal L} Q_r)}
	{(\nu \seq n)(\sigma~|~(\nu_{j \in\mathbb N\backslash\mathcal I}{\alphac_j}_?{\betac_j}_?{\deltac_j}_?{\alpha_j}_?{\beta_j}_?{\delta_j}_?)~P'')~~\mathcal V~~(\nu \seq n)(\sigma~|~Q'')}
$$
}}
\caption{Simulation relation for Lemma \ref{dynamic-safetylemma} ($\psi[\phi[\_]] \preccurlyeq \_$)}
\label{fig:simreln-4-d}
\end{figure}

We prove that the relations $\mathcal S$, $\mathcal T$, and $\mathcal U$ in Figures \ref{fig:simreln-1-d}, \ref{fig:simreln-2-d}, and \ref{fig:simreln-3-d} are included in the simulation preorder. Some interesting points in those proofs are listed below. 
\begin{itemize}
\item In Section \ref{sec:mpstat}, when an operation request is sent in ${\it TS}^s$ we send an appropriate authorization request in ${\it NAS}^s$, obtain a capability, and send an execution request with that capability (see $\mathcal T$ in Figure \ref{fig:simreln-2-s}). In contrast, here when an operation request is sent in ${\it TS}^d$ we \emph{wait} after sending an appropriate authorization request in ${\it NAS}^d$ (see $\mathcal T$ in Figure \ref{fig:simreln-2-d}); we continue only when that operation request in ${\it TS}^d$ is processed, when we obtain a capability in ${\it NAS}^d$, send an execution request with that capability, and process the execution request. 

But why wait? Suppose that the operation request in ${\it TS}^d$ carries a time bound $\infty$; now if we obtain a capability in ${\it NAS}^d$ before the operation request in ${\it TS}^d$ is processed, we commit to a finite time bound, which breaks the simulation. 
\item As before, $\phi[\psi]$ forces a fresh capability to be acquired for every execution request by filtering execution requests in ${\it NAS}^d$ through ${\it TS}^d$ and back. When an execution request is sent in ${\it NAS}^d$ under $\phi[\psi]$ we  send an execution request with the same capability in ${\it NAS}^d$ (see $\mathcal U$ in Figure \ref{fig:simreln-3-d}). But under $\phi[\psi]$ a fresh capability is obtained and the execution request is sent again with the fresh capability. If the capability in the original request expires before the fresh capability, the simulation breaks. Fortunately operation requests in ${\it TS}^d$ carry time bounds, so we can communicate this expiry bound through ${\it TS}^d$. In fact there seems to be no way around this problem \emph{unless} time bounds can be specified in operation requests in ${\it TS}^d$!
\end{itemize}
By Proposition \ref{pf-prec} we have:
\begin{lemma}\label{dynamic-lemma} For any $F$, $\rho$, and $C$,
\begin{enumerate}
\item $(\nu_{i \in \mathcal I}\alpha_i\beta_i\delta_i)(C~|~\new{K_{M\!D}K'_M}\textsc{nafs}(F,\varnothing,0,\rho)) $
\par $~~~~~~~~~~~~~~~~~~~~~~~~~~~~~~~~~~~~~~~~~~~~~~~~~~~~~~~\preceq~ \phi[(\nu_{i \in \mathcal I}\alphac_i\betac_i\deltac_i)(\lceil C\rceil~|~\textsc{tfs}(\lceil F\rceil,\varnothing,0,\lceil \rho\rceil))] $
\item $(\nu_{i \in \mathcal I}\alphac_i\betac_i\deltac_i)(\lceil C\rceil~|~\textsc{tfs}(\lceil F\rceil,\varnothing,0,\lceil \rho\rceil))$
\par $~~~~~~~~~~~~~~~~~~~~~~~~~~~~~~~~~~~~~~~~~~~~~~~~~~\preceq ~ \psi[(\nu_{i \in \mathcal I}\alpha_i\beta_i\delta_i)(C~|~\new{K_{M\!D}K'_M}\textsc{nafs}(F,\varnothing,0,\rho))] $
\item $\phi[\psi[(\nu_{i \in \mathcal I}\alpha_i\beta_i\delta_i)(C~|~\new{K_{M\!D}K'_M}\textsc{nafs}(F,\varnothing,0,\rho))]] $
\par $~~~~~~~~~~~~~~~~~~~~~~~~~~~~~~~~~~~~~~~~~~~~~~~~~~~~~~~\preceq ~ (\nu_{i \in \mathcal I}\alpha_i\beta_i\delta_i)(C~|~\new{K_{M\!D}K'_M}\textsc{nafs}(F,\varnothing,0,\rho))$
\end{enumerate}
\end{lemma}
So by Proposition \ref{pf-fullabs}, $\mathcal R$ is secure. 

Further we prove that the relation $\mathcal V$ in Figure \ref{fig:simreln-4-d} is also included in the simulation preorder. By Proposition \ref{pf-prec} we have: 
\begin{lemma}\label{dynamic-safetylemma} For any $F$, $\rho$, and $C$, 
\vspace{0.2cm}
\par $\psi[\phi[(\nu_{i \in \mathcal I}\alphac_i\betac_i\deltac_i)(\lceil C\rceil~|~\textsc{tfs}(\lceil F\rceil,\varnothing,0,\lceil \rho\rceil))]] $
\par $~~~~~~~~~~~~~~~~~~~~~~~~~~~~~~~~~~~~~~~~~~~~~~~~~~~~~~~~~\preceq ~ (\nu_{i \in \mathcal I}\alphac_i\betac_i\deltac_i)(\lceil C\rceil~|~\textsc{tfs}(\lceil F\rceil,\varnothing,0,\lceil \rho\rceil))$
\end{lemma}
So by Lemmas \ref{dynamic-lemma}.1--2 and Corollary \ref{pf-pres}, $\mathcal R$ is safe and fully abstract. 

%\section{More counterexamples?}
%The implementation-specification relation does not expose the same public interfaces. In cases where they do, we could say more.

%In the case where the relation exposes the public interface of the specification, we may not have any more counterexamples. However in the case where the public interface of the implementation is exposed instead, \emph{i.e.}, ${\it NAS}$ and $\bowtie\!\textsc I[\:\emph{IS}\:]$ are related, we readily have counterexamples whenever we have counterexamples to
%$$\bowtie\!\textsc I[\:\bowtie\! \textsc{Na}[\:{\it NAS}\:]\:] \not\preceq {\it NAS}$$
%Indeed we can prove
%$${\shortuparrow^\mathrm{NAS}_\mathrm{TS}}[\:\bowtie\! \textsc I[\:\emph{IS}\:]\:] \preceq \emph{IS}$$
%and thus
%$$\bowtie\! \textsc I[\:{\shortuparrow^\mathrm{NAS}_\mathrm{TS}}[\:\bowtie\! \textsc I[\:\emph{IS}\:]\:]\:] \preceq \bowtie\! \textsc I[\:\emph{IS}\:]$$
%This class of counterexamples involves testing with arbitrary evaluation contexts.

%\end{document}

\section{Designing secure distributed protocols}\label{apply}
\noindent
In the preceding sections, we present a thorough analysis of the problem of distributing access control. Let us now apply that analysis to a more general problem.
% that of distributing an arbitrary stateful computation. 

Suppose that we are required to design a distributed protocol that securely implements a specification. (The specification may be an arbitrary computation.) We can solve this problem by partitioning the specification into smaller computations, running those computations in parallel, and securing the intermediate outputs of those computations so that they may be released and absorbed in any order. 
In particular, we can design ${\it NS}^{d+}$ by partitioning ${\it IS}^{d+}$ into access control and storage, running them in parallel, and securing the intermediate outputs of access control as capabilities. % (so that they may be released and absorbed in any order). 
The same principles should guide any such design. For instance,  by (R\ref{Rfake}) and (R\ref{Rdiff}) intermediate outputs should not leak information prematurely;  by (R\ref{Rexpire}) and (R\ref{Rdelay}) such outputs must be timestamped and the states on which they depend must not change between clock ticks; and  by (A\ref{Abound}) the specification must be generalized with time bounds.% in the specification.

\paragraph{\em Computation as a graph}
\noindent
We describe a computation as a directed graph $G\cal (V,E)$. % with maximum outdegree $1$ (a forest). %where $\{\max(d)~|~d = \mathtt{out}(v), ~v \in V\} = 1$. 
The \emph{input nodes}, collected by $\mathcal V_i \subseteq \mathcal V$, are the nodes of indegree $0$. The \emph{output nodes}, collected by $\mathcal V_o \subseteq \mathcal V$, are the nodes of outdegree $0$. Further, we consider a set of  \emph{state nodes} $\mathcal V_s \subseteq \mathcal V$ such that $\mathcal V_i \cap \mathcal V_s = \varnothing$. As a technicality, any node that is in a cycle or has outdegree $> 1$ must be in $\mathcal V_s$. 
%We assume that every state node has a self-edge, that is, $E \supseteq E_s$ where $E_s = \{(v,v)~|~v \in V_s\}$. We also assume that $E \setminus E_s$ is acyclic. 

Nodes other than the input nodes run some code. Let $\mathcal M$ contain all terms and $\sqsubset$ be a strict total order on $\mathcal V$. We label each $v \in \mathcal V \setminus (\mathcal V_i \cup \mathcal V_s)$ with a function $\lambda_v : \mathcal M^{\In(v)} \rightarrow \mathcal M$, and each $v \in \mathcal V_s$ with a function $\lambda_v : \mathcal M^{\In(v)}\times \mathcal M \rightarrow \mathcal M$. %For any such node, each incoming edge draws an argument, and for state nodes, the state is an additional argument. %These functions model code run at the internal nodes. 
Further, each state node carries a shared clock, following the midnight-shift scheme. 

A \emph{configuration} $(\pisymbol,\tau)$ consists of a partial function $\pisymbol : \mathcal V \rightarrow \mathcal M$ such that $\dom(\pisymbol) \supseteq \mathcal V_s$, and a total function $\tau : \mathcal V_s \rightarrow \mathbb N$. Intuitively, $\pisymbol$ assigns values at the state nodes and some other nodes, and $\tau$ assigns times at the state nodes. For any $v \in \mathcal V \setminus \mathcal V_i$, the function $\lambda_v$ outputs the value at $v$, taking as inputs the values at each incoming $u$, and the value at $v$ if $v$ is a state node; further, if such $u \notin \mathcal V_s$, the value at $u$ is ``consumed" on input. Formally, the operational semantics is given by a binary relation $\rightsquigarrow$ over configurations.
% modulo an arbitrary total ordering $\sqsubseteq$ on $V$.
%\pagebreak
$$\infer{v \in \mathcal V \setminus (\mathcal V_i \cup \mathcal V_s) \\ \forall k \in 1..\In(v).~(u_k,v) \in \mathcal E~\wedge~\pisymbol(u_k) = t_k  \\  u_1 \sqsubset \dots \sqsubset u_{\In(v)} \\ \pisymbol^- = \pisymbol |_{\mathcal V_s \cup (\mathcal V \setminus \{u_1,\dots,u_{\In(v)}\}}}
{(\pisymbol,\tau) \rightsquigarrow (\pisymbol^- [v \mapsto \lambda_v(t_1,\dots,t_{\In(v)})],\tau)}$$
$$\infer{v \in \mathcal V_s \\ \tau(v) = \clk \\ \pisymbol(v) = t 
 \\\\ \forall k \in 1..\In(v).~(u_k,v) \in \mathcal E~\wedge~\pisymbol(u_k) = t_k   \\\\ u_1 \sqsubset \dots \sqsubset u_{\In(v)} \\ \pisymbol^- = \pisymbol |_{\mathcal V_s \cup (\mathcal V \setminus \{u_1,\dots,u_{\In(v)}\}}}
{(\pisymbol,\tau) \rightsquigarrow (\pisymbol^- [v \mapsto \lambda_v(t_1,\dots,t_{\In(v)},t)],\tau[v \mapsto \clk + 1])}$$
As usual, we leave the context implicit; the adversary is an arbitrary context that can write values at $\mathcal V_i$, read values at $\mathcal V_o$, and read times at $\mathcal V_s$.

For example, a graph that describes ${\it IS}^{d+}$ is:
%\footnote{The graph is approximate because we do not yet have clocks.}
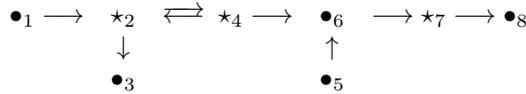
\[
\left.
\begin{array}{lcccr}
\bullet_1 \longrightarrow & \star_2 & \longrightleftarrows \star_4 \longrightarrow & \bullet_6 & \longrightarrow \star_7 \longrightarrow \bullet_8\\
& \downarrow & & \uparrow\\
& \bullet_3 & & \bullet_5
\end{array}
\right.
\]
Here $\mathcal V_i = \{\bullet_1,\bullet_5\}$, $\mathcal V_o = \{\bullet_3,\bullet_8\}$, $\mathcal V_s = \{\star_2,\star_4,\star_7\}$, and $\mathcal V = \mathcal V_i \cup \mathcal V_o \cup \mathcal V_s \cup \{\bullet_6\}$. Intuitively, $\star_2$ carries accumulators, and $\bullet_1$ and $\bullet_3$ carry inputs and outputs for access modifications; $\star_4$ carries access policies, and $\bullet_6$ carries access decisions; $\star_7$ carries stores, and $\bullet_5$ and $\bullet_8$ carry inputs and outputs for store operations. We define: %; given such a configuration, computation proceeds as follows:
\begin{eqnarray*}
%\item $\bullet_1$ gets $\tup{k,\op}$ and $\bullet_6$ gets $\tup{k',\op'}$.
\lambda_{\star_2}(\tup{k,\theta},F,\tup{\_,\Xi}) & = & \func{exec}(\mathtt{perm}_{F,k,\theta},\theta,\Xi) \\
\lambda_{\bullet_3}(\tup{N,\Xi}) & = & N \\
\lambda_{\star_4}(\tup{\_,\Xi},\_) & = & \Xi \\
\lambda_{\bullet_6}(F,\tup{k,\op}) & = & \tup{\op,\mathtt{perm}_{F,k,\op}} \\
\lambda_{\star_7}(\tup{\op,L},\tup{\_,\rho}) & = & \func{exec}(L,\op,\rho) \\
\lambda_{\bullet_8}(\tup{N,\rho}) & = & N 
%\item $\bullet_6$ and $\bullet_6$ may plug into other computation graphs.
\end{eqnarray*}
%\renewcommand{\theenumi}{\arabic{enumi}}
%\renewcommand{\labelenumi}{\theenumi.}
%
%$\pisymbol(\star_2) = \tup{\_,F}$ and $\pisymbol(\star_7) = \tup{\_,\rho}$ for some $F$ and $\rho$; $\tau(\star_2) = \tau(\star_7) = 0$; and
%
%(The careful reader may note that this model is not entirely accurate: $\star_2$ should carry accumulators, and a state node should be introduced between $\star_2$ and $\bullet_6$ to carry access policies. We ignore this minor detail for simplicity.)

\paragraph{\em Distribution as a graph cut}
\noindent
Once described as a graph, a computation can be distributed along any cut of that graph. %\footnote{A cut is a set of edges that disconnects a graph.} 
For instance, ${\it IS}^{d+}$ can be distributed along the cut $\{(\bullet_6,\star_7)\}$ to obtain ${\it NS}^{d+}$. We present this derivation formally in several steps. 
%For convenience, we assume that the specification $G(\mathcal V,\mathcal E)$ is written in a special form, which we describe next.

%\vspace{-2mm}
\paragraph{Step 1} 
For each $v \in \mathcal V$, let $S(v) \subseteq \mathcal V_s$ be the set of state nodes that have paths to $v$, and $I(v) \subseteq \mathcal V_i$ be the set of input nodes that have paths to $v$ without passing through nodes in $\mathcal V_s$.   Then $G(\mathcal V,\mathcal E)$ can be written in a form where, loosely, the values at $I(v)$ and the times at $S(v)$ are explicit in $\pisymbol(v)$ for each node $v$.  
Formally, the \emph{explication} of $G$ is the graph $\hat G(\hat \mathcal V,\hat \mathcal E)$ where $\hat \mathcal V = \mathcal V \cup \{\hat v~|~v \in \mathcal V_i\} \cup \{\hat u~|~u \in \mathcal V_o\}$ and $\hat \mathcal E = \mathcal E \cup  \{(\hat v,v)~|~v \in \mathcal V_i\} \cup \{(u,\hat u)~|~u \in \mathcal V_o\}$. We define:
$$\infer{v \in \mathcal V_i}
{\hat \lambda_v(t) = \tup{t,t}} 
\qquad
\infer{v \in \mathcal V_o}
{\hat \lambda_{\hat v}(\_,t) = t}$$
$$\infer{v \in \mathcal V\setminus (\mathcal V_i \cup \mathcal V_s) \\ \lambda_v(t_1,\dots,t_{\In(v)}) = t}
{\hat \lambda_v(\tup{I_1,t_1}, \dots, \tup{I_{\In(v)},t_{\In(v)}}) = \tup{\tup{I_1\dots I_{\In(v)}}, t}}$$
$$\infer{v \in \mathcal V_s \\ \pisymbol(v) = \tup{\clk,t} \\ \lambda_v(t_1,\dots,t_{\In(v)}, t) = t'}
{\hat \lambda_v(\tup{\_,t_1}, \dots, \tup{\_,t_{\In(v)}}, \tup{\clk,t}) = \tup{\clk+1,t'}}$$
This translation is sound and complete. 
\begin{theorem}\label{thm:closure} $\hat G$ is fully abstract with respect to $G$.
\end{theorem}
\noindent
For example, the explication of the graph for ${\it IS}^{d+}$ is:%\footnote{The graph is approximate because we do not yet have clocks.}
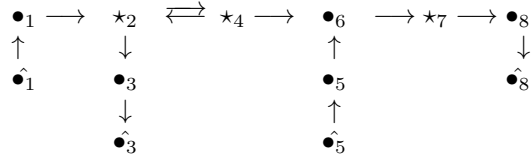
\[
\left.
\begin{array}{lcccr}
\bullet_1 \longrightarrow & \star_2 & \longrightleftarrows \star_4 \longrightarrow & \bullet_6 & \longrightarrow \star_7 \longrightarrow \bullet_8\\
\uparrow &  \downarrow & & \uparrow & \downarrow\\
\hat{\bullet_1} & \bullet_3 & & \bullet_5 & \hat{\bullet_8} \\
& \downarrow & & \uparrow\\
& \hat{\bullet_3} & & \hat{\bullet_5}
\end{array}
\right.
\]
Here $\pisymbol(\bullet_6)$ is of the form $\tup{\tup{k,\op,\clk},\tup{\op,\mathtt{perm}_{F,k,\op}}}$ rather than $\tup{\op,\mathtt{perm}_{F,k,\op}}$;  the ``input" $\pisymbol(\hat{\bullet_5}) = \tup{k,\op}$, the ``time" $\tau(\star_4) = \clk$, and the ``output" $\tup{\op,\mathtt{perm}_{F,k,\op}}$ of an access check are all explicit in $\pisymbol(\bullet_6)$. A capability can be conveniently constructed from this form (see below). 

%\vspace{-2mm}
\paragraph{Step 2}
%By Theorem \ref{thm:closure}, suppose that $G(\mathcal V,\mathcal E)$ is the explicit form of some graph that describes a computation. 
Next, let $\mathcal E_0$ be any cut. As a technicality, we assume that $\mathcal E_0 \cap ((\mathcal V_i \cup \mathcal V_s) \times \mathcal V) = \varnothing$. The \emph{distribution} of $G$ along $\mathcal E_0$ is the graph $ G^\$(\mathcal V^\$,\mathcal E^\$)$, where $\mathcal V^\$ = \hat\mathcal V \cup \{ \overline v~|~(v,\_) \in \mathcal E_0\} \cup \{ v^\$~|~(v,\_) \in \mathcal E_0\}$ and $\mathcal E^\$ = (\hat \mathcal E \setminus \mathcal E_0)  \cup \{(\overline v,v^\$)~|~(v,\_) \in \mathcal E_0\} \cup \{(v^\$, u)~|~(v,u) \in \mathcal E_0\}$. 
Let $K_v$ and $E_v$ be secret keys shared by $v$ and $v^\$$ for every $(v,\_) \in \mathcal E_0$. We define: 
$$\infer{(v,\_) \in \mathcal E_0 \\ \hat \lambda_v(t_1,\dots,t_{\In(v)}) = \tup{t,t'} \\ m\mbox{ is fresh}}
{ \lambda^\$_v(t_1,\dots,t_{\In(v)}) = \func{mac}(\tup{t,\{m,t'\}_{E_v}},K_v)}$$
$$\infer{(v,\_) \in \mathcal E_0 \\ \tau(S(v))\mbox{ is included in }t}
{ \lambda^\$_{v^\$}(\tup{t,\func{mac}(\tup{t,\{\_,t'\}_{E_v}},K_v)}) = \tup{t,t'}}$$
$$\infer{v \in \mathcal V \setminus \mathcal V_i \\ (v,\_) \notin \mathcal E_0}{ \lambda^\$_v = \hat \lambda_v}$$
\noindent
Intuitively, for every $(v,\_) \in \mathcal E_0$, $v^\$$ carries the same values in $G^\$$ as $v$ does in $G$; those values are encoded and released at $v$, absorbed at $\overline v$, and decoded back at $v^\$$. For example, the distribution of the graph for ${\it IS}^{d+}$ along the cut $\{(\bullet_6,\star_7)\}$ is:
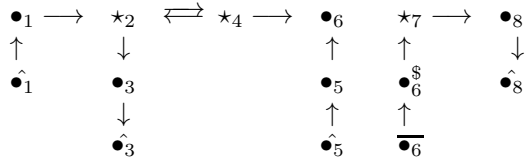
\[
\left.
\begin{array}{lcccclr}
 \bullet_1 \longrightarrow & \star_2&  \longrightleftarrows \star_4 \longrightarrow & \bullet_6 & & \star_7 \longrightarrow & \bullet_8\\
\uparrow &\downarrow & & \uparrow && \uparrow & \downarrow\\
\hat{\bullet_1} &\bullet_3 & & \bullet_5 && \bullet_6^\$ & \hat{\bullet_8}  \\
&\downarrow & & \uparrow && \uparrow \\
& \hat{\bullet_3} & & \hat{\bullet_5} &&\overline{\bullet_6}
\end{array}
\right.
\]
%\[
%\left.
%\begin{array}{ccr}
% \longrightarrow  \longrightarrow &  \\
%&& \downarrow \\
%&& 
%\end{array}
%\right.
%\]
This graph describes a variant of ${\it NS}^{d+}$. In particular, 
the node $\bullet_6$ now carries a capability of the form $\func{mac}(\tup{\tup{k,\op,\clk},\{m,\tup{\op,\mathtt{perm}_{F,k,\op}}\}_{E_{\bullet_6}}},K_{\bullet_6})$, that secures the input, time, and output of an access check.

%\vspace{-2mm}
\paragraph{Step 3} Finally, $G$ is revised following (A\ref{Abound}). The \emph{revision} of $G$ along $\mathcal E_0$ is the graph $G^\#(\mathcal V^\#,\mathcal E^\#)$, where $\mathcal V^\# = \mathcal V \cup \{v^\#~|~(v,\_) \in \mathcal E_0\}$ and $\mathcal E^\# = \mathcal E \cup \{(v^\#,v)~|~(v,\_) \in \mathcal E_0\}$. We define:
%
%\vspace{-2mm}
$$\infer{(v,u) \in \mathcal E_0 \\ \tau(S(v)) \leq T}
{\lambda^\#_v(t_1,\dots,t_{\In(v)},T) = \lambda_v(t_1,\dots,t_{\In(v)})}$$
$$\infer{v \in \mathcal V \setminus \mathcal V_i \\ (v,\_) \notin \mathcal E_0}{\lambda^\#_v = \lambda_v}$$
Intuitively, for every $(v,\_) \in \mathcal E_0$, progress at $v$ requires that the times at $S(v)$ do not exceed the time bounds at $v^\#$. For example, the revised form of the graph for ${\it IS}^{d+}$ is: 
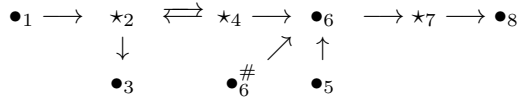
\[
\left.
\begin{array}{lcrcr}
\bullet_1 \longrightarrow & \star_2 & \longrightleftarrows \star_4 \longrightarrow \!\!& \bullet_6 & \longrightarrow \star_7 \longrightarrow \bullet_8\\
& \downarrow & \nearrow\!\! & \uparrow\\
& \bullet_3 & \bullet_6^\#~~~~& \bullet_5
\end{array}
\right.
\]
%Here $\lambda^\#_{\bullet_6}$ behaves as $\lambda_{\bullet_6}$ after taking as input a time bound $\pisymbol(\bullet_6^\#)$ and checking that it is equal to $\tau(\star_4)$.
Here $\bullet_6^\#$ carries a time bound $T$, and $\lambda^\#_{\bullet_6}(F,\tup{k,\op},T) = \lambda_{\bullet_6}(F,\tup{k,\op})$ if $\tau(\star_4) \leq T$. 

We prove the following correctness result.
\begin{theorem}\label{thm:general} $G^\$$ is fully abstract with respect to $G^\#$.
\end{theorem}
\noindent
By Theorem \ref{thm:general}, the graph for ${\it NS}^{d+}$ is fully abstract with respect to the revised graph for ${\it IS}^{d+}$. 

%For the simpler case of static access policies, consider the subgraph derived from ${\it IS}^{d+}$ by removing the nodes $\{\bullet_1,\star_2,\bullet_3\}$; this graph describes ${\it IS}^s$.
%\[
%\left.
%\begin{array}{lr}
% \bullet_6 & \longrightarrow \star_7 \longrightarrow \bullet_8 \\
% \uparrow  \\
% \bullet_5
%\end{array}
%\right.
%\]
%We define $\lambda_{\bullet_6}(\tup{k,\op}) = \tup{\op,\mathtt{perm}_{F,k,\op}}$ for some static $F$. Next, distributing along the cut $\{(\bullet_6,\star_7)\}$, we have:
%\[
%\left.
%\begin{array}{cclr}
%\bullet_6 & & \star_7 \longrightarrow & \bullet_8\\
%\uparrow && \uparrow & \downarrow\\
%\bullet_5 && \ddot{\bullet_6} & \hat{\bullet_8}  \\
%\uparrow && \uparrow \\
%\hat{\bullet_5} &&(\bullet_6,\star_7)
%\end{array}
%\right.
%\]
%This graph describes a variant of ${\it NS}^s$. Now $\pisymbol(\bullet_6)$ is of the form $\func{mac}(\tup{\tup{k,\op},\{m,\tup{\op,\mathtt{perm}_{F,k,\op}}\}_{E_{\bullet_6}}},K_{\bullet_6})$---in particular, capabilities do not carry timestamps. This graph is fully abstract to a trivial revision of ${\it IS}^s$:
%\[
%\left.
%\begin{array}{rr}
% \bullet_6 & \longrightarrow \star_7 \longrightarrow \bullet_8 \\
%\nearrow~ \uparrow  \\
% \bullet_6^\#~~~~~~\bullet_5
%\end{array}
%\right.
%\]
Similarly, we can design ${\it NS}^s$ from ${\it IS}^s$. The induced subgraph of ${\it IS}^{d+}$ without $\{\bullet_1,\star_2,\bullet_3,\star_4\}$ describes ${\it IS}^s$. We define $\lambda_{\bullet_6}(\tup{k,\op}) = \tup{\op,\mathtt{perm}_{F,k,\op}}$ for some static $F$. Distributing along the cut $\{(\bullet_6,\star_7)\}$, we obtain the induced subgraph of ${\it NS}^{d+}$ without $\{\hat\bullet_1,\bullet_1,\star_2,\bullet_3,\hat\bullet_3,\star_4\}$. 
This graph describes a variant of ${\it NS}^s$, with $\pisymbol(\bullet_6)$ of the form $\func{mac}(\tup{\tup{k,\op},\{m,\tup{\op,\mathtt{perm}_{F,k,\op}}\}_{E_{\bullet_6}}},K_{\bullet_6})$. (Here capabilities do not carry timestamps.) By Theorem \ref{thm:general}, the graph for ${\it NS}^s$ is fully abstract with respect to a trivially revised graph for ${\it IS}^s$, where $\lambda^\#_{\bullet_6}(\tup{k,\op},\tup{}) = \lambda_{\bullet_6}(\tup{k,\op})$.

\section{Conclusion}
\noindent
We present a comprehensive analysis of the problem of implementing distributed access control with capabilities. In previous work, we show how to implement static access policies securely \cite{ChaudhuriAbadi-FMSE05} and dynamic access policies safely \cite{ForteChaudhuriA06}. In this paper, we explain those results in new light, revealing the several pitfalls that any such design must care about for correctness, while discovering interesting special cases that allow simpler implementations. Further, we present new insights on the difficulty of implementing dynamic access policies securely (a problem that has hitherto remained unsolved). We show that such an implementation is in fact possible if the specification is slightly generalized. 

Moreover, our analysis turns out to be surprisingly general. Guided by the same basic principles, we show how to automatically derive secure distributed implementations of other stateful computations. %This connection seems to be more than a coincidence. 
This approach is reminiscent of secure program partitioning \cite{spp}, and investigating its scope should be interesting future work.

\paragraph{Acknowledgments}
This work owes much to Mart\'in Abadi, who formulated the original problem and co-authored our previous work in this area. Many thanks to him and Sergio Maffeis for helpful discussions on this work, and detailed comments on an earlier draft of this paper. It was Mart\'in who suggested the name ``midnight-shift". Thanks also to him and C\'edric Fournet for clarifying an issue about the applied pi calculus, which led to simpler proofs.

\bibliography{ref}
\bibliographystyle{abbrv} %alpha

\end{document}